\documentclass[11pt]{article}

\usepackage{epsfig,subfigure}
\usepackage{amssymb}
\usepackage[fleqn]{amsmath}
\usepackage{color}
\usepackage{arydshln}

\usepackage{graphicx}
\usepackage{color}
\usepackage[dvipsnames]{xcolor}

\definecolor{armygreen}{rgb}{0.29, 0.33, 0.13}

\usepackage{amssymb}
\usepackage{amsmath,amsfonts,amssymb}
\usepackage{verbatim}
\usepackage{stfloats}
\usepackage[bookmarks=false]{}
\usepackage{algorithm}
\usepackage{algcompatible}
\usepackage{dsfont}
\usepackage{mathtools}
\usepackage{enumitem}
\usepackage{multirow}
\newtheorem{theorem}{Theorem}

\newtheorem{lemma}{Lemma}

\newtheorem{remark}{Remark}
\newtheorem{observation}{Observation}

\newcommand\blfootnote[1]{%
  \begingroup
  \renewcommand\thefootnote{}\footnote{#1}%
  \addtocounter{footnote}{-1}%
  \endgroup
}

\newcommand{\calE}{\mathcal{E}}

\newcommand{\calI}{\mathcal{I}}
\newcommand{\calJ}{\mathcal{J}}
\newcommand{\calK}{\mathcal{K}}
\newcommand{\calL}{\mathcal{L}}

\newcommand{\calT}{\mathcal{T}}

\newcommand{\bfC}{\mathbf{C}}

\newcommand{\bfF}{\mathbf{F}}
\newcommand{\bfG}{\mathbf{G}}
\newcommand{\bfH}{\mathbf{H}}
\newcommand{\bfI}{\mathbf{I}}

\newcommand{\bfM}{\mathbf{M}}
\newcommand{\bfN}{\mathbf{N}}

\newcommand{\bfP}{\mathbf{P}}

\newcommand{\bfU}{\mathbf{U}}
\newcommand{\bfV}{\mathbf{V}}

\newcommand{\F}{\mathbb{F}}
\newcommand{\Fq}{\mathbb{F}_{q}}

\newcommand{\smallo}{\ensuremath{o}}

\newcommand\bigZero{\mbox{\LARGE$\mathbf{0}$}}
\newcommand\bigM{\mbox{\LARGE$\mathbf{M}$}}
\newcommand\bigI{\mbox{\LARGE$\mathbf{I}$}}

\newcommand{\layer}{\iota}
\newcommand{\Minter}{\bfM_{\textrm{interference}} }
\newcommand{\nodeset}{\mathcal{N}}
\newcommand{\queries}{\mathcal{Q}}
\newcommand{\secrecy}{S}

\newcommand{\findex}{\kappa}
\newcommand{\Neff}{N_{\text{eff}}}

\mathchardef\mhyphen="2D 

\begin{document}
\date{}
\title{
The Capacity of Private Information Retrieval with Eavesdroppers
}
\author{\normalsize Qiwen Wang$^{*}$, Hua Sun$^{\dagger}$, Mikael Skoglund$^{*}$ \\
           {\small $^{*}$Department of Information Science and Engineering, KTH Royal Institute of Technology} \\
           {\small $^{\dagger}$Department of Electrical Engineering, University of North Texas} \\
           {\small \it Email: \{qiwenw, skoglund\}@kth.se, \{hua.sun\}@unt.edu }
}

\maketitle

\begin{abstract}
We consider the problem of private information retrieval (PIR) with colluding servers and eavesdroppers (abbreviated as ETPIR). 
The ETPIR problem is comprised of $K$ messages, $N$ servers where each server stores all $K$ messages, a user who wants to retrieve one of the $K$ messages without revealing the desired message index to any set of $T$ colluding servers, and an eavesdropper who can listen to the queries and answers of any $E$ servers but is prevented from learning any information about the messages. 
The information theoretic capacity of ETPIR is defined to be the maximum number of desired message symbols retrieved privately per information symbol downloaded. 
We show that the capacity of ETPIR is $C = \left( 1- \frac{E}{N} \right)  \left(1 +  \frac{T-E}{N-E} + \cdots + \left( \frac{T-E}{N-E} \right)^{K-1} \right)^{-1}$ when $E < T$, and $C = \left( 1 - \frac{E}{N} \right)$ when $E \geq T$.
To achieve the capacity, the servers need to share a common random variable (independent of the messages), and its size must be at least $\frac{E}{N} \cdot \frac{1}{C}$ symbols per message symbol. 
Otherwise, with less amount of shared common randomness,
ETPIR is not feasible and the capacity reduces 
to zero.

\vspace{0.1in}
An interesting observation is that the ETPIR capacity expression takes different forms in two regimes. When $E < T$, the capacity equals the inverse of a sum of a geometric series with $K$ terms and decreases with $K$; this form is typical for capacity expressions of PIR. When $E \geq T$, the capacity does not depend on $K$, a typical form for capacity expressions of SPIR (symmetric PIR, which further requires data-privacy, {\it i.e.,} the user learns no information about other undesired messages); the capacity does not depend on $T$ either.
In addition, the ETPIR capacity result 
includes multiple previous PIR and SPIR capacity results as special cases.\blfootnote{This paper was presented in part at Asilomar 2018.} 


\end{abstract}

\newpage
\section{Introduction}
The problem of private information retrieval (PIR) is to retrieve one message out of $K$ messages privately from $N$ servers (each stores all $K$ messages) without letting any server know the identity of the requested message. An important factor in the design of such PIR protocols is the communication efficiency. 
Recently, a series of works studies diverse variations of the PIR problem from an information-theoretic perspective. Among these works, the messages are considered to be long sequences, such that the upload cost is neglected as it is diminishing when compared to the download cost. The communication efficiency is measured by the capacity, which is defined as the maximum number of desired information bits retrieved privately per information bit downloaded.

The first work in this series is~\cite{sun2017capacity}, which finds the PIR capacity as $C_{\text{PIR}} = \left(  1 + \frac{1}{N} + \cdots + \left( \frac{1}{N} \right)^{K-1}\right)^{-1}$. 
A significant number of following works add various elements to the PIR problem to form different variations. Among these elements, three are most relevant to this work: colluding servers, adversaries, and symmetric PIR (SPIR). 

We first discuss related works on colluding servers. In the modeling of the initial work~\cite{sun2017capacity}, the servers do not share information about their communication with the user, {\it i.e.,} they do not collude to infer the identity of the requested message. However, it is possible in practical systems that some sets of servers can communicate and then potentially collude. 
In~\cite{sun2017colluding}, the problem where any set of $T$ servers may collude 
is studied (TPIR) and the capacity is established to be $C_{\text{TPIR}} = \left(  1 + \frac{T}{N} + \cdots + \left( \frac{T}{N} \right)^{K-1}\right)^{-1}$. 
Beyond TPIR, PIR with arbitrary collusion patterns (where the cardinality of each set of colluding servers may vary) is first studied in \cite{tajeddine2017private} and the capacity of one particular case is found in \cite{jia2017capacity}, where the colluding sets are disjoint.

The second element is adversaries, referring to scenarios where the communication system is vulnerable to passive and/or active attacks.
This is also a practical factor to consider in PIR protocol design, because applications of cloud storage are usually built on open systems and networks, where preserving data security and privacy is crucial and more challenging.
The problem of TPIR with an active adversary who may erase the responses from any $A$ servers is considered in \cite{sun2017colluding} and interestingly, the capacity in this setting equals the capacity of TPIR with $N-A$ servers (the number of surviving servers). The problem of TPIR with a Byzantine (active) adversary who may introduce arbitrary errors to the responses from any $B$ servers is considered in \cite{banawan2017capacity} and interestingly, the capacity in this setting equals the capacity of TPIR with $N-2B$ servers times $(N-2B)/N$ (where $2B$ out of all $N$ servers are used to identify and correct the errors and are thus wasted, akin to the Hamming distance requirement in algebraic coding theory).
In this work, we consider the problem of TPIR with a passive eavesdropper who may hear the responses from any $E$ servers while learning nothing about the messages (ETPIR). The ETPIR problem is first studied in~\cite{wang2017EPIR,wang2018BEpir}, where an achievable scheme and a converse are provided, although the two do not match.  We will close the gap and establish the exact capacity in this work. It turns out that neither the achievable scheme nor the converse in~\cite{wang2017EPIR,wang2018BEpir} is tight and we need to improve both.
{\color{black}
Recent work~\cite{banawan2018pir_wiretap} considers a related PIR setting with eavesdroppers, but the eavesdropping model is different. In~\cite{banawan2018pir_wiretap}, the eavesdropper can wiretap all servers, but can only observe a fraction of each server's answer. While in our work, the eavesdropper can only wiretap $E$ (instead of all $N$) servers, but if a server is wiretapped, then its answer is fully (instead of partially) observed by the eavesdropper.
}

We finally consider SPIR.
Relative to PIR, SPIR requires one more constraint on data-privacy at the user side. Namely, the user should not learn any information about the other messages besides the desired one.
Although this work focuses exclusively on the PIR problem, 
our result turns out to relate intimately to 
SPIR so that a brief summary of known results along this line is presented next. For a PIR problem, if the letter `$S$' is added to its description, we are referring to the same problem with SPIR constraint. For example, SPIR refers to the PIR problem with data-privacy requirement, and its capacity is shown to be $C_{\text{SPIR}}=1-\frac{1}{N}$ \cite{sun2016symmetric}.
The capacity of TSPIR is characterized as $C_{\text{TSPIR}} = 1-\frac{T}{N}$, by specializing a more general result from \cite{wang2017linear}.
The capacity of TSPIR with Byzantine (active) adversaries 
is characterized as $C_{\text{BTSPIR}} = 1-\frac{2B+T}{N}$ in~\cite{wang2017secure,wang2018BEpir}.
The capacity of ETSPIR is established to be $C_{\text{ETSPIR}}=1-\frac{\max(T,E)}{N}$ in~\cite{wang2017secure,wang2018BEpir}.
An interesting observation from these results is that the capacity of a PIR problem is always the inverse of a sum of a geometric series with $K$ terms, and decreases with the number of messages $K$ while the capacity of an SPIR problem does not depend $K$. Furthermore, the SPIR capacity is the limit of the corresponding PIR capacity by taking $K \to \infty$. 

\begin{figure}[t] 
    \centering
        \includegraphics[width=0.90\textwidth]{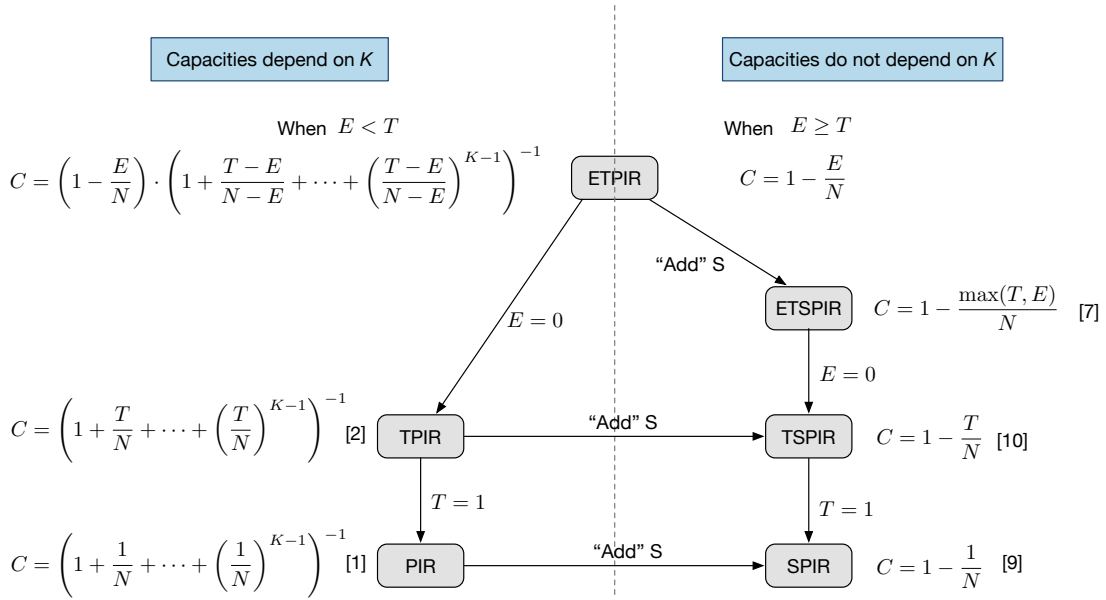}
    \caption{Connections to prior works and results.} \label{fig:intro_graph}
\end{figure}

The main contribution of this work is that we characterize the capacity of ETPIR (where two elements -- eavesdropper and colluding servers are added to the conventional setting of PIR). 
Specifically, $K$ messages are replicated at $N$ servers. A user wants to retrieve one message by communicating with the servers, without revealing the identity of the desired message. Any $T$ servers may collude, that is, they may share their communication with the user to infer the desired message index. The system is vulnerable to eavesdroppers, who are curious about the messages and can tap in on the communication between any $E$ servers and the user. The ETPIR protocol should prevent the eavesdropper from learning any information about the messages.
We show that the capacity of ETPIR is
\begin{eqnarray}
C_{\text{ETPIR}} = 
\begin{cases}
\left( 1 - \frac{E}{N} \right) \cdot \left( 1 + \frac{T-E}{N-E} +  \left( \frac{T-E}{N-E} \right)^2 + \cdots +  \left( \frac{T-E}{N-E} \right)^{K-1} \right)^{-1}, & \text{when } E < T; \\
1-\frac{E}{N}, & \text{when } E \geq T.
\end{cases} 
\end{eqnarray}
To achieve the capacity, the servers need to share some common randomness that is independent of the messages, in the amount of at least an $\frac{E}{N}$ fraction of one over the capacity 
times the message size. Otherwise, the capacity is zero and the ETPIR problem is not feasible.

The ETPIR capacity result includes the capacity results of various previously studied settings as special cases, e.g., TPIR, PIR, ETSPIR, TSPIR and SPIR. A pictorial illustration is shown in Figure~\ref{fig:intro_graph}.
It is interesting that in the regime of $E \geq T$, 
ETPIR capacity is always equal to ETSPIR capacity (instead of approaching it in the large $K$ limit), {\it i.e.,} there is no penalty in the retrieval rate by further requiring data-privacy at the user side. 

\subsection{Notations} \label{sec:notations}
Let $[m:n]$ denote the set $\{m, m+1, \dots, n\}$ for $m \leq n$, and let $(m:n)$ denote the vector $(m, m+1, \dots, n)$. To simplify the notation, denote the set of random variables $\{ X_m, X_{m+1}, \dots, X_n\}$ by $X_{[m:n]}$ and the vector $( X_m, X_{m+1}, \dots, X_n )$ by $X_{(m:n)}$.
For an index set $\calI = \{ i_1, i_2, \dots, i_n \}$, denote the set of variables 
$\{ X_i: i \in \calI \}$ by $X_{\calI}$.
The cardinality of a set $\mathcal{A}$ is denoted by $|\mathcal{A}|$.

We denote the dimension of a matrix by the superscript 
(the dimension is omitted when there is no ambiguity). For example, a matrix $\bfM$ with $m$ rows and $n$ columns is denoted by $\bfM^{m \times n}$. 
When the matrix is square ($m=n$), the notation is simplified to $\bfM^m$.
The identity matrix with dimension $m \times m$ is denoted by $\bfI^m$. 
{\color{black} The set of all full-rank $n \times n$ matrices over a finite field $\Fq$ is denoted by $GL_n(\Fq)$.}
For a matrix $\bfM$, $\bfM[\calI,:]$ denotes the submatrix of $\bfM$ formed by the rows corresponding to the elements of the vector $\calI$.
If $\bfG$ is comprised of $A \times B$ square matrices with dimension $I \times I$, {\it i.e.,} $\bfG$ is a block matrix with dimensions $AI \times BI$, 
\begin{eqnarray}
\bfG = 
\begin{bmatrix}
\bfM_{1,1} & \cdots & \bfM_{1,B} \\
\vdots & \ddots & \vdots \\
\bfM_{A,1} & \cdots & \bfM_{A,B}
\end{bmatrix},
\end{eqnarray}
where ${\bf M}_{i,j}$ are all $I \times I$ matrices,
denote the $i$-th \emph{block row} by $\bfG^{I}[i,:]$ ({\it i.e.}, the submatrix of $\bfG$ comprised of $I$ rows from the $[(i-1)I+1]$-th row to the $iI$-th row),
\begin{eqnarray}
\bfG^{I}[i,:] = 
\begin{bmatrix}
\bfM_{i,1} & \cdots & \bfM_{i,B}
\end{bmatrix}.
\end{eqnarray}
\emph{Block column} is defined in a similar way.
The notation $\otimes $ denotes the Kronecker product. 
We denote the $m \times n$ Vandermonde matrix generated from $n$ distinct symbols $\lambda_1, \lambda_2, \dots, \lambda_n$ from a finite field by $\bfV^m(\lambda_1,  \dots, \lambda_n)$, {\it i.e.,}
\begin{eqnarray}
\bfV^m(\lambda_1,  \dots, \lambda_n) = 
\begin{bmatrix}
1 & 1 & \cdots & 1 \\
\lambda_1 &  \lambda_2 & \cdots &  \lambda_n \\
\vdots & \vdots & \ddots & \vdots \\
\lambda_1^{m-1} &  \lambda_2^{m-1} & \cdots &  \lambda_n^{m-1}
\end{bmatrix}.
\end{eqnarray}

\section{Problem Setup}
A database comprised of $K$ messages is replicated at $N$ servers. The messages $\{ W_k \}$ are independent and each message consists of $L$ symbols from $\Fq$. \begin{eqnarray}
H(W_k) = L , \text{for } k=1,\dots,K  \quad ; \quad H(W_1, \dots, W_K) = KL.
\end{eqnarray}
Here and throughout the paper we measure entropy to base $q$.

A user wants to retrieve a message $W_{\findex}$ with index $\findex$ from the servers, where the desired message index $\findex$ follows some prior distribution over $[1:K]$. 
{\color{black} Denote the realization of $\findex$ by $k$. Based on $k$, the user generates random queries to send to the servers.}
The query received by Server $n$ is denoted by $Q_{n}^{[k]}$. Let $\queries = \left[Q_{n}^{[k]} \right]_{n \in [1:N], k \in [1:K]}$ denote the complete query scheme, namely, the collection of all queries under all choices of the desired message index.
{\color{black} The queries are independent of the messages, {\it i.e.,}}
\begin{eqnarray}
I(\queries ; W_{[1:K]}) = 0. \label{eqn:UQindepW}
\end{eqnarray}

In the communication system, an eavesdropper is interested in the messages and can tap in on the communication between any $E$ servers and the user. To prevent the eavesdropper from learning the messages, the servers share a common random variable, denoted by $S$, which is independent of the messages and the queries, {\it i.e.,}
\begin{eqnarray}
I(S; W_{[1:K]}, \queries) = 0. \label{eqn:SindepWQ}
\end{eqnarray}
Let $\rho$ denote the ratio of the amount of common randomness relative to the message size, {\it i.e.,}
\begin{eqnarray}
\rho \triangleq \frac{H(S)}{H(W_k)} = \frac{H(S)}{L}.
\end{eqnarray}

The servers follow the protocol agreed with the user {\it a priori}, and generate answers based on the received query $Q_n^{[k]}$, the stored messages $W_{[1:K]}$, and the common random variable $S$. The answer sent to the user from Server $n$ is denoted by $A_{n}^{[k]}$. We have
\begin{eqnarray}
H(A_n^{[k]} | Q_n^{[k]}, W_{[1:K]}, S) = 0.
\end{eqnarray}

The eavesdropper must learn no information about the messages from the queries and answers of any $E$ servers. That is, the following $E$-security constraint must be satisfied,
\begin{eqnarray}
\text{[$E$-security]} \qquad I(W_{[1:K]} ; A_{\calE}^{[k]}, Q_{\calE}^{[k]}) = 0, \forall \calE \subset [1:N], |\calE| = E. \label{eqn:modelsecurity}
\end{eqnarray}

Any $T$ servers may collude. To guarantee user-privacy, from the queries and answers of any $T$ servers, together with the message contents and the common random variable, the servers should not be able to infer any information about the desired message index. Thus, the following $T$-privacy constraint must be satisfied,
\begin{eqnarray}
\text{[$T$-privacy]} \qquad I(A_{\calT}^{[\findex]}, Q_{\calT}^{[\findex]} , W_{[1:K]}, S ; \findex)=0, \forall \calT \subset [1:N], |\calT| = T. \label{eqn:modelprivacy}
\end{eqnarray}

From all the answers downloaded from the servers and the available information at the user, the user should be able to decode the desired message with diminishing probability of error as $L$ tends to infinity. By Fano's inequality, this corresponds to the following correctness constraint,
\begin{eqnarray}
\text{[Correctness]} \qquad H(W_k | A_{[1:N]}^{[k]}, Q_{[1:N]}^{[k]}) = \smallo(L). \label{eqn:modelcorrect}
\end{eqnarray}

The ETPIR rate, $R$ of a scheme characterizes the number of desired information symbols retrieved per downloaded information symbol, 
\begin{eqnarray}
R = \frac{H(W_k)}{\sum_{n=1}^{N} H(A_n^{[k]})} = \frac{L}{\sum_{n=1}^{N} H(A_n^{[k]})}.
\end{eqnarray}
A rate $R$ is said to be $\epsilon$-error achievable if there exists a sequence of ETPIR schemes with rate at least $R$, and probability of error $P_e \to 0$ as $L \to \infty$. The supremum of all $\epsilon$-error achievable rates is called the $\epsilon$-error capacity $C$.

\section{Main Result}
Theorem~\ref{thm:capacity} below states the main result of this work.
\begin{theorem} \label{thm:capacity}
The capacity of the ETPIR problem is
\begin{eqnarray}
C_{\text{ETPIR}} = 
\begin{cases}
\left( 1 - \frac{E}{N} \right) \cdot \left( 1 + \frac{T-E}{N-E} +  \left( \frac{T-E}{N-E} \right)^2 + \cdots +  \left( \frac{T-E}{N-E} \right)^{K-1} \right)^{-1}, & \text{when } E < T; \\
1-\frac{E}{N}, & \text{when } E \geq T.
\end{cases} \label{eqn:theorem}
\end{eqnarray}
To achieve the capacity, 
$\rho_{\text{ETPIR}} \geq \frac{E}{N} \cdot C_{\text{ETPIR}}^{-1}$. If $\rho_{\text{ETPIR}} < \frac{E}{N} \cdot C_{\text{ETPIR}}^{-1}$, 
the ETPIR problem is not feasible.
\end{theorem}

For our theorem, we have the following interesting observations.
\begin{observation}
When $E=0$, $C_{\text{ETPIR}} = C_{\text{TPIR}} = \left(1+\frac{T}{N}+\cdots + (\frac{T}{N})^{K-1} \right)^{-1}$. Therefore, Theorem \ref{thm:capacity} includes the capacity of TPIR~\cite{sun2017colluding} as a special case.
\end{observation}

\begin{observation}
When $T=1$, $C_{\text{ETPIR}} = C_{\text{EPIR}} = 1 - \frac{E}{N} $. Note that the EPIR problem (PIR with eavesdroppers) has not been studied in the literature. Therefore, the capacity of EPIR is obtained from Theorem \ref{thm:capacity} as a rewarding by-product.
\end{observation}

\begin{observation}
When $E=0$, $T=1$, $C_{\text{ETPIR}} = C_{\text{PIR}} = \left(1+\frac{1}{N}+\cdots + (\frac{1}{N})^{K-1} \right)^{-1}$. Therefore, Theorem \ref{thm:capacity} includes the capacity of PIR~\cite{sun2017capacity} as a special case.
\end{observation}

\begin{observation}
{\color{black} When $T \leq N-E$, our achievable scheme attains the capacity with zero-error
. 
When $T > N-E$, our achievable scheme has $\epsilon$-error. 
It is interesting to see if the scheme can be strengthened to have zero-error.}
\end{observation}

\begin{observation}
When $E<T$, the capacity is the product of two terms. The first term $1-\frac{E}{N} = \frac{N-E}{N}$ has an {\color{black} intuitive} illustration -- from the security constraint~\eqref{eqn:modelsecurity}, the answers from any $E$ servers provide no useful information to decode the messages (out of the $N$ servers, only $N-E$ are useful. Thus we have the ratio $\frac{N-E}{N}$). 
The second term $\left( 1 + \frac{T-E}{N-E} +  \left( \frac{T-E}{N-E} \right)^2 + \cdots +  \left( \frac{T-E}{N-E} \right)^{K-1} \right)^{-1}$ is more intriguing. Note that it is equal to the capacity of TPIR with $N-E$ servers where any $T-E$ servers might collude. 
If we view the first $E$ servers as noise (common randomness) providers, then we are left with $N-E$ servers that return noiseless answers. For the remaining $N-E$ servers, a first thought is that any $T$ (instead of $T-E$) servers might collude. Interestingly, it turns out that the first $E$ noise providing servers also contribute to reduce the $T$-privacy constraint in the original $N$ servers setting to the $(T-E)$-privacy constraint in the $N-E$ servers setting after noise is cancelled.
\end{observation}

\begin{observation}
When $E \geq T$, the second term disappears (the $T$-privacy constraint is dominated by the $E$-security constraint) and the capacity depends neither on $K$ nor on $T$. 
As the capacity does not depend on $T$, there is no penalty in increasing $T$ to $E$, {\it i.e.,} we automatically have $E$-privacy (see the achievable scheme in Section~\ref{sec:schemeEgreaterT}). 
\end{observation}

\begin{observation}
When $E \geq T$, the capacity of ETPIR turns out to be equal to 
the capacity of \mbox{ETSPIR}~\cite{wang2017secure}. 
As ETSPIR requires one more data-privacy constraint at the user, it is evident that the converse of ETPIR also holds for ETSPIR.  Our achievable scheme for ETPIR directly guarantees data-privacy such that 
data-privacy is obtained for free. 
\end{observation}

\begin{observation}
When $E < T$, $\lim_{K \to \infty} C_{\text{ETPIR}} =  C_{\text{ETSPIR}}$ (consistent with previous relations between PIR capacity and SPIR capacity); when $E \geq T$, $C_{\text{ETPIR}} \equiv  C_{\text{ETSPIR}}$ (distinct from previous relations). 
\end{observation}

\section{Achievability when $E<T$} \label{sec:schemeEsmallerT}
In this section, we present the proof of the achievability part of Theorem~\ref{thm:capacity} for the case of $E<T$. 
In the setting of TPIR~\cite{sun2017colluding}, the queries are constructed using an MDS code. However, for our setting of ETPIR (with an additional element of eavesdroppers), MDS coded queries no longer suffice and we need a more sophisticated design. A high level description of the protocol design is as follows.
As the eavesdropper can tap in on any $E$ answers (meaning that the combination of any $E$ answers can not contain any useful information about the messages), the shared common randomness is coded with an $(N,E)$-MDS matrix and added to the answers as the noise to guarantee $E$-security and the property that the noise can be cancelled by projecting the answers to the {\color{black} dual} space of the $(N,E)$-MDS code. 
After the noise is cancelled, we wish to ensure that the undesired symbols retain certain linear dependency (for example, MDS property suffices) such that they (named {\it interference}) can be cancelled and the desired symbols can be decoded. Beyond this correctness constraint, we further need to satisfy $T$-privacy by ensuring that every $T$ servers observe linearly independent queries. For such a purpose, an $(N,T)$-MDS structure on the queries is desired. To summarize, the essence of our scheme is that the queries are coded by a (first) MDS code, and after projecting queries to the {\color{black} dual} space of the (second) MDS code of the common randomness, the projected queries still form a (third) MDS code. 
Next we illustrate this design progressively by a series of examples with increasing complexity in Sections~\ref{sec:achieve_ex1} - \ref{sec:achieve_ex5}, leading to the general scheme in Section~\ref{sec:achievable_general}. 

\subsection{Elemental case $K=2, N=3, T=2, E=1$} \label{sec:achieve_ex1}
Let's start with the smallest setting with $K=2$ messages, $N=3$ servers where any $T=2$ may collude, and an eavesdropper who may hear the communication associated with any $E=1$ server. 


Let $\{ a_i \}$ and $\{ b_i \}$ denote the {\it mixtures} (linear combinations) of symbols from $W_1$ and $W_2$ respectively, and let $r, s, t$ denote three uniform independent symbols shared by the servers (unknown to the eavesdropper and the user). It is natural to start with a scheme in the following form where the answer from 1 server is pure noise. Without loss of generality, suppose the user wants to retrieve $W_1$. 

\begin{center}
\begin{tabular}{|c|c|c|}
\hline
\small{Server 1} & \small{Server 2} & \small{Server 3} \\
\hline
$a_1 + r$           & $a_3 + r$           & $ r$ \\
$b_1 + s$          & $b_2 + s$           & $ s$ \\
$a_2 + b_2 + t$ & $a_4 + b_1 + t$ & $ t$ \\
\hline
\end{tabular}
\end{center}

Although the above scheme is correct as noise variables and non-desired symbols can be cancelled, it is not $2$-private as Server 1 and Server 2 can collude and figure out that the $b$ symbols have the same indices while the $a$ symbols have distinct indices. 
To make it 2-private, we add message symbols with carefully chosen linear coefficients to Server 3 (so that it will enhance privacy and simultaneously act as the noise provider), as shown below. 

\begin{center}
\begin{tabular}{|c|c|c|}
\hline
\small{Server 1} & \small{Server 2} & \small{Server 3} \\
\hline
$a_1 + r$           & $a_3 + r$           & $a_5 + r$ \\
$b_1 + s$          & $b_3 + s$           & $b_5 + s$ \\
$a_2 + b_2 + t$ & $a_4 + b_4 + t$ & $a_6 + b_6 + t$ \\
\hline
\end{tabular}
\end{center}

The specific construction of the mixtures $\{ a_i \}$ and $\{ b_i \}$ is as follows. Assume each message consists of 4 symbols from a field $\Fq$ with $q \geq 3$. 
Let $W_1$ and $W_2$ denote the column vectors comprised of the $L = 4$ symbols of each message.
The user privately chooses two $4 \times 4$ matrices $\bfU_1$ and $\bfU_2$ independently and uniformly from {\color{black} $GL_4(\Fq)$}. 
Recall that $a_{(1:4)}$ and $b_{(1:4)}$ denote the column vectors comprised of $\{a_1, \dots, a_4 \}$ and $\{ b_1, \dots, b_4 \}$, respectively. They are generated by
\begin{eqnarray}
a_{(1:4)} & = &  \bfU_1 W_1, \\
b_{(1:4)} & = &  \bfU_2 W_2.
\end{eqnarray}
%

Further, $a_5, a_6, b_5, b_6$ are assigned as follows.
\begin{center}
\begin{tabular}{|c|c|c|}
\hline
\small{Server 1} & \small{Server 2} & \small{Server 3} \\
\hline
$a_1 + r$           & $a_3 + r$           & $a_1+ a_3 + r$ \\
$b_1 + s$          & $b_3 + s$           & $b_3 + ( b_4-b_2) + s$ \\
$a_2 + b_2 + t$ & $a_4 + b_4 + t$ & $(a_2+ a_4 )+ ( b_3 -b_1)+ (2b_4 -b_2)  + t$ \\
\hline
\end{tabular}
\end{center}
{\color{black}
The query sent to Server 1 is $(\bfU_1[1,:], \bfU_2[1,:], \bfU_1[2, :], \bfU_2[2,;])$, i.e., the coefficients for the message symbols. Using the coefficients received, the servers then produce the answers according to the table above. 
}

Next, we remove the noise variables by subtracting the answers of Server 1 from the answers of Server 2 and Server 3.
\begin{center}
\begin{tabular}{|c|c|}
\hline
 \small{Server 2 $-$ Server 1} & \small{Server 3 $-$ Server 1} \\
\hline
 $(a_3 - a_1)$                        &         $a_3$   \\
 $(b_3 - b_1)$                        &         $(b_3 - b_1) + (b_4 -b_2)   $ \\
$(a_4 - a_2) + (b_4 - b_2)$ & $a_4  + (b_3 - b_1) + 2(b_4 -b_2)  $ \\
\hline
\end{tabular}
\end{center}

We are now ready to prove that the scheme satisfies the required properties.

\noindent {\bf Correctness:} We call one row of the retrieval table a \emph{query row}. That is, a query row is comprised of one symbol (in the same row) downloaded from each server.
After removing the noise symbols $r, s, t$, from the first query row ({\it i.e.,} $a_1+r, a_3+r, a_1+a_3+r$), the user decodes $a_1$ and $a_3$. From the second query row, the user obtains $b_3 - b_1$ and $b_4 - b_2$, which are used to cancel the interference caused by $W_2$ in the third query row. After canceling the interference, the user decodes $a_2$ and $a_4$ from the third query row.
Because $\bfU_1$ is full-rank and known by the user, the user can decode the 4 symbols of $W_1$ from $a_{(1:4)}$.

\noindent {\bf 1-security:} 
Because $r,s,t$ are uniformly distributed random variables independent of the message symbols, we have $I(W_1, W_2 ; A_n^{[k]}, Q_n^{[k]}) = I(W_1, W_2 ; A_n^{[k]} | Q_n^{[k]}) 
= H(A_n^{[k]} | Q_n^{[k]}) - H(A_n^{[k]}|W_1, W_2, Q_n^{[k]}) 
= H(A_n^{[k]} | Q_n^{[k]}) - H(r, s,t) = 0$. Therefore, from the answer of any server, the eavesdropper can learn nothing about the messages $W_1, W_2$.


\noindent {\bf $2$-privacy:} Privacy is guaranteed by the observation that the answers of any 2 servers contain 
4 linearly independent random combinations of symbols from each message. 
For example, Server 1 and Server 2 observe $a_{(1:4)}$ and $b_{(1:4)}$, which are linearly independent random combinations of the 2 messages. Server 1 and Server 3 observe  
\begin{eqnarray}
\left[
\begin{matrix}
a_1 ,  a_2  , b_1 , b_2 \\
a_1+a_3 , a_2+a_4 , -b_2+b_3+b_4 ,  -b_1-b_2+b_3+2b_4
\end{matrix}
\right]
\xLeftrightarrow{\text{invertible}}
\left[
\begin{matrix}
a_1 , a_2  , b_1 , b_2\\
a_3 , a_4  , b_3 , b_4
\end{matrix} 
\right] \label{eqn:ex1privacy}
\end{eqnarray}
Therefore, Server 1 and Server 3 also observe linearly independent random combinations of $W_1$ and $W_2$. The case where Server 2 and Server 3 collude follows similarly.

Because $\bfU_1$ and $\bfU_2$ are independently and uniformly chosen from all $4 \times 4$ full-rank matrices, the mapping from $W_1$ to $a_{(1:4)}$ is i.i.d. as the mapping from $W_2$ to $b_{(1:4)}$. The invertible relation in~\eqref{eqn:ex1privacy} indicates that the random variables on both sides of \eqref{eqn:ex1privacy} are identically distributed (a more detailed proof of this argument appears in Lemma 1 of~\cite{sun2017colluding}), so that 
any 2 servers cannot distinguish $W_1$ and $W_2$ and the scheme is private. 

\noindent {\bf Achievable rate:} The scheme downloads 9 symbols in total, out of which the user can decode 4 desired symbols. Therefore, the rate achieved by this scheme is $4/9$, which matches the capacity promised in Theorem~\ref{thm:capacity}. The achieved common randomness size $\rho$ ($1/3$ of the total download) is also as promised.

In this section, we present a capacity achieving scheme for the smallest non-trivial setting. When we proceed to larger parameters, 
the main challenge is to design the linear combination coefficients of the desired and undesired symbols (named \emph{precoding} or the \emph{precoding matrices}), such that
\begin{enumerate}
\item Any $T$ servers observe linearly independent combinations of symbols from each message (to guarantee $T$-privacy);
\item After the noise is cancelled, the interference of undesired symbols can be canceled as well;
\item After the noise and the interference of undesired symbols are cancelled, the desired symbols have full rank (to guarantee that the desired message can be recovered). 
\end{enumerate}

\subsection{More layers when the number of messages $K$ increases, $K=3, N=3, T=2, E=1$} \label{sec:achieve_ex2}
We consider an example where the number of messages increases. In this case, the scheme has more layers and interestingly, the layers can be designed independently. 

Suppose each message consists of 8 symbols from a finite field $\Fq$ with $q \geq 3$. The user privately chooses 3 matrices $\bfU_1, \bfU_2, \bfU_3$ uniformly and independently from {\color{black} $GL_8(\Fq)$.} 
Let $a_{(1:8)} = \bfU_1 W_1$, $b_{(1:8)} = \bfU_2 W_2$, and $c_{(1:8)} = \bfU_3 W_3$, and let $s_{(1:7)}$ contain the noise variables. The retrieval scheme for $W_1$ is shown below, which includes 3 layers separated by dashed lines. 
\begin{center}
\begin{tabular}{|c|c|c|}
\hline
\small{Server 1} & \small{Server 2} & \small{Server 3} \\
\hline
$a_1 + s_1$           & $a_5 + s_1$           & $a_1 + a_5 + s_1$ \\
$b_1 + s_2$          & $b_5 + s_2$           & $b_5+ (b_6-b_2) + s_2$ \\
$c_1 + s_3$          & $c_5 + s_3$           & $c_5+ (c_6-c_2) + s_3$ \\
\hdashline
$a_2 + b_2 + s_4$     & $a_6 + b_6 + s_4$     & $a_2+a_{6}+ (b_5-b_1) + (2b_{6}-b_2)+ s_4$ \\
$a_3 + c_2 + s_5$     & $a_7 + c_6 + s_5$     & $a_3+a_{7}+ (c_5-c_1)+ (2c_{6} -c_2) + s_5$ \\
$b_3 + c_3 + s_6$     & $b_7 + c_7 + s_6$     & $b_7+c_7 + (b_8+ c_{8}-b_4-c_4) + s_6$ \\
\hdashline
$a_{4} + b_{4} + c_{4} +s_7$     & $a_{8} + b_{8} + c_{8} +s_7$     & $a_4+a_{8} + (b_{7}+c_7-b_3-c_3)+(2b_8+ 2c_{8}-b_4-c_4)  +s_7$ \\
\hline
\end{tabular}
\end{center}

Note that the scheme uses the same precoding in Section~\ref{sec:achieve_ex1} 3 times, {\it i.e.,} for the $b$ symbols in layer 1 and layer 2, the $c$ symbols in layer 1 and layer 2, and the $b+c$ symbols in layer 2 and layer 3. The design cross different layers may not be the same in general (for more details, refer to Section~\ref{sec:achievable_general}).

Security is obvious. Correctness and privacy can be checked similar to the proof in Section~\ref{sec:achieve_ex1}.
The rate achieved is $8/21$, matching the capacity in Theorem~\ref{thm:capacity}.

\subsection{Precoding design when the number of servers $N$ increases, $K=2, N=4, T=2, E=1$}  \label{sec:achieve_ex3}
Assume each message consists of 9 symbols  from a field $\Fq$ with $q \geq 9$ and the user wants to retrieve $W_1$. In the following retrieval table, 
the 4 symbols $r, s, t, z$ are independent and uniform random symbols privately shared by the servers.
\begin{center}
\begin{tabular}{|c|c|c|c|}
\hline
\small{Server 1} & \small{Server 2} & \small{Server 3} & \small{Server 4} \\
\hline
$a_1 + r$           & $a_4 + r$           & $a_7 + r$  &       $a_{10} + r$ \\
$b_1 + s$           & $b_4 + s$           & $b_7 + s$  &       $b_{10} + s$ \\
$a_2 + b_2 + t$     & $a_5 + b_5 + t$     & $a_{8} + b_{8}+ t$    & $a_{11} + b_{11}+ t$ \\
$a_3 + b_3 + z$     & $a_6 + b_6 + z$     & $a_{9} + b_{9}+ z$    & $a_{12} + b_{12}+ z$ \\
\hline
\end{tabular}
\end{center}

The user privately chooses 2 $9 \times 9$ matrices $\bfU_1$ and $\bfU_2$ from {\color{black} $GL_9(\Fq)$.} 
Let $\bfG_1 = \bfV^4(\phi_1, \phi_2, \phi_3)$, {\it i.e.,} a Vandermonde matrix with distinct nonzero $\phi_i$'s, and $\phi_i \neq 1, i \in [1:3]$,
\begin{eqnarray}
\bfG_1 = 
\begin{bmatrix}
1 & 1 & 1 \\
\phi_1 & \phi_2 & \phi_3 \\
\phi_1^2 & \phi_2^2 & \phi_3^2 \\
\phi_1^3 & \phi_2^3 & \phi_3^3
\end{bmatrix}. \label{eqn:ex3G1}
\end{eqnarray}
Let $\bfG_2$ denote a $12 \times 9$ matrix, 
\begin{eqnarray}
\bfG_2 = 
\begin{bmatrix}
\bfI^3 & \mathbf{0} \\
\mathbf{0} & \bfI^3 \\
\bfI^3 - \bfM_1 & \bfM_1 \\
\bfI^3 - \bfM_2 & \bfM_2
\end{bmatrix}, \label{eqn:ex3G2}
\end{eqnarray}
where $\bfI^3$ is the $3 \times 3$ identity matrix, and $\bfM_1$ and $\bfM_2$ are $3 \times 3$ matrices specified as follows. For 3 distinct nonzero elements $\psi_1, \psi_2, \psi_3$ that are not the 6-th roots of unity in $\Fq$, {\it i.e.,} $\psi_i^6 \neq 1, i \in [1:3]$, we set
\begin{eqnarray}
\bfM_1 = \bfV^3(\psi_1, \psi_2, \psi_3) \cdot \text{diag}(\psi_1^3, \psi_2^3, \psi_3^3) \cdot [\bfV^3(\psi_1, \psi_2, \psi_3)]^{-1}, \label{eqn:ex3M1} \\
\bfM_2 = \bfV^3(\psi_1, \psi_2, \psi_3) \cdot \text{diag}(\psi_1^6, \psi_2^6, \psi_3^6) \cdot [\bfV^3(\psi_1, \psi_2, \psi_3)]^{-1}. \label{eqn:ex3M2}
\end{eqnarray}

$a_{(1:12)}$ are designed as follows, where each query row is precoded by $\bfG_1$.
\begin{eqnarray}
[a_1, a_4, a_7, a_{10}]^T =  \bfG_1  \bfU_1[(1:3),:] W_1,  \label{eqn:ex3desiredprecode1} \\
{ [a_2, a_5, a_8, a_{11}]^T } =  \bfG_1  \bfU_1[(4:6),:] W_1,  \label{eqn:ex3desiredprecode2}  \\
{ [a_3, a_6, a_9, a_{11}]^T } =  \bfG_1  \bfU_1[(7:9),:] W_1,  \label{eqn:ex3desiredprecode3} 
\end{eqnarray}
%

$b_{(1:12)}$ are precoded by $\bfG_2$,
\begin{eqnarray}
b_{(1:12)} = \bfG_2 \bfU_2[(1:6), :] W_2. \label{eqn:ex3undesiredprecode}
\end{eqnarray}

The description of the scheme is complete and we prove that the scheme is correct, private, secure and capacity achieving.

\noindent {\bf Correctness:}
We first cancel the noise.
\begin{center}
\begin{tabular}{|c|c|c|}
\hline
 \small{Server 2 $-$ Server 1} & \small{Server 3 $-$ Server 1} & \small{Server 4 $-$ Server 1} \\
\hline
 $a_4 - a_1 $           & $a_7 - a_1$  &       $a_{10} - a_1$ \\
 $b_4 - b_1 $           & $b_7 - b_1 $  &       $b_{10} - b_1 $ \\
 $(a_5 - a_2) + (b_5 - b_2)$     & $(a_{8}-a_2) + (b_{8}-b_2)$    & $ (a_{11}-a_2) + (b_{11}-b_2)$ \\
 $(a_6-a_3) + (b_6-b_3) $     & $(a_{9}-a_3) + (b_{9}-b_3)$    & $(a_{12}-a_3) + (b_{12} - b_3)$ \\
\hline
\end{tabular}
\end{center}
From the precoding of $b_{(1:12)}$~\eqref{eqn:ex3undesiredprecode} and the structure of $\bfG_2$~\eqref{eqn:ex3G2}, for the undesired symbols we have
\begin{eqnarray}
\begin{bmatrix}
b_7 - b_1 \\
 b_8 - b_2 \\
 b_9 - b_3 \\
 b_{10}-b_1 \\
 b_{11}-b_2 \\
 b_{12}-b_3
\end{bmatrix}
=
\begin{bmatrix}
\bfM_1 \\
\bfM_2 
\end{bmatrix}
\cdot 
\begin{bmatrix}
b_4-b_1 \\
b_5-b_2 \\
 b_6-b_3
\end{bmatrix}.
\end{eqnarray}

From the definition of $\bfM_1$~\eqref{eqn:ex3M1} and $\bfM_2$~\eqref{eqn:ex3M2}, we know that $[\bfI^3; \bfM_1; \bfM_2]$
$
= \bfV^9(\psi_1, \psi_2, \psi_3) \cdot [\bfV^3(\psi_1, \psi_2, \psi_3)]^{-1}$
is the systematic form of a Vandermonde matrix, and thus is a $9 \times 3$ MDS matrix. Therefore, the 3 exposed symbols $b_4-b_1, b_7-b_1, b_{10}-b_1$ in the second query row are sufficient to reconstruct the remaining 6 symbols $b_5-b_2, b_6-b_3, b_8-b_2, b_9-b_3, b_{11}-b_2, b_{12}-b_3$ in the third and fourth query rows. Hence, the interference caused by $b_{(1:12)}$ can be cancelled and the user obtains
\begin{center}
\begin{tabular}{|c|c|c|}
\hline
\small{Server 2 $-$ Server 1} & \small{Server 3 $-$ Server 1} & \small{Server 4 $-$ Server 1} \\
\hline
 $a_4 - a_1 $           & $a_7 - a_1$  &       $a_{10} - a_1$ \\
 $a_5 - a_2 $     & $a_{8}-a_2 $    & $ a_{11}-a_2 $ \\
 $a_6-a_3  $     & $a_{9}-a_3 $    & $a_{12}-a_3 $ \\
\hline
\end{tabular}
\end{center}
From the definition of $\bfG_1$~\eqref{eqn:ex3G1}, and the assumption that the $\phi_i$'s are nonzero, distinct, and $\phi_i \neq 1$, it is easy to check that each query row above contains 3 linear independent combinations of the desired symbols. Further, ${\bf U}_1$ has full rank and the user can invert the 9 linear independent  combinations of the desired symbols to decode $W_1$.


\noindent {\bf $2$-privacy:}
From the precoding of the desired symbols~\eqref{eqn:ex3desiredprecode1}-\eqref{eqn:ex3desiredprecode3}, the linear combinations of $W_1$ across query rows are independent. In addition, from the structure of $\bfG_1$~\eqref{eqn:ex3G1}, within each query row, every 3 $a_i$ symbols are linearly independent. Therefore, every 2 servers observe 6 linear independent combinations of $W_1$.

We next show that every 2 servers observe linear independent combinations of $W_2$. From~\eqref{eqn:ex3undesiredprecode}, $\bfG_2$ expands 6 linear independent combinations of $W_2$ to 12 symbols, which are distributed to the servers. The precoding coefficients of the 6 $b_i$ symbols observed by any 2 servers are given by the following $6$ matrices (each of which is a submatrix of ${\bf G}_2$),
\begin{eqnarray}
\begin{bmatrix}
\bfI^3 & \mathbf{0} \\
\mathbf{0} & \bfI^3
\end{bmatrix},
\begin{bmatrix}
\bfI^3 & \mathbf{0} \\
\bfI^3 - \bfM_1 & \bfM_1
\end{bmatrix},
\begin{bmatrix}
\bfI^3 & \mathbf{0} \\
\bfI^3 - \bfM_2 & \bfM_2
\end{bmatrix},
\begin{bmatrix}
\mathbf{0} & \bfI^3 \\
\bfI^3 - \bfM_1 & \bfM_1
\end{bmatrix},
\begin{bmatrix}
\mathbf{0} & \bfI^3 \\
\bfI^3 - \bfM_2 & \bfM_2
\end{bmatrix},
\begin{bmatrix}
\bfI^3 - \bfM_1 & \bfM_1 \\
\bfI^3 - \bfM_2 & \bfM_2
\end{bmatrix}.
\end{eqnarray}

We are left to prove that these 6 matrices are invertible, which is
is equivalent to prove that  $\bfM_1$, $\bfM_2$, $\bfI^3-\bfM_1$, $\bfI^3-\bfM_2$ and 
$\bfM_1-\bfM_2$
are invertible. This follows from~\eqref{eqn:ex3M1},~\eqref{eqn:ex3M2} and the assumption that $\psi_i \neq 0$, $\psi_i^6 \neq 1$. 

Therefore, any 2 servers observe linear independent random combinations of $W_1$ and $W_2$ so that the mappings to $W_1$ and $W_2$ are i.i.d. and 2-privacy is guaranteed.

Security is easy to verify. The rate achieved is equal to the capacity, $9/16$ and the size of common randomness used is $1/4$ of the total download, as desired.

\subsection{Precoding of noise symbols when $E > 1$, $K=2, N=5, T=3, E=2$}  \label{sec:achieve_ex4}
In the 3 examples above, $E=1$ and all servers use the same noise (common randomness) symbols to add on the message symbols ({\it i.e.,} repetition code is used).
When $E > 1$, the noise symbols are precoded by an $(N,E)$-MDS code. 
To illustrate how this is done, we present an example with $E = 2$. 

Assume each message is comprised of 9 symbols from a sufficiently large field $\Fq$, and the user wants $W_1$. In the scheme described below, the user downloads 20 symbols in total, achieving the capacity of $9/20$.
\begin{center}
{\renewcommand{\arraystretch}{1.1}
\begin{tabular}{|c|c|c|c|c|}
\hline
\small{Server 1} & \small{Server 2} & \small{Server 3} & \small{Server 4} & \small{Server 5} \\
\hline
$a_1 + r_1$           & $a_4 + r_2$           & $a_7 + r_1+ r_2$  &       $a_{10} +  r_1 +2r_2$ & $a_{13} + 2 r_1 +3r_2$ \\
$b_1 + s_1$           & $b_4 +s_2$           & $b_7 + s_1+s_2$  &       $b_{10} + s_1+2s_2$  & $b_{13} + 2s_1+3s_2$ \\
$a_2 + b_2 + t_1$     & $a_5 + b_5 + t_2$     & $a_{8} + b_{8}+ t_1+t_2$    & $a_{11} + b_{11}+ t_1+2t_2$ & $a_{14} + b_{14}+ 2t_1+3t_2$ \\
$a_3 + b_3 + z_1$     & $a_6 + b_6 + z_2$     & $a_{9} + b_{9}+ z_1+z_2$    & $a_{12} + b_{12}+ z_1+2z_2$  & $a_{15} + b_{15}+ 2z_1+3z_2$ \\
\hline
\end{tabular}
}
\end{center}

In the retrieval table above, the 8 shared symbols $r_1, r_2, s_1, s_2, t_1,t_2, z_1, z_2$ are uniform and independent random variables. 
The precoding matrix and parity check matrix of the noise symbols are
\begin{eqnarray}
\bfC_S = 
\begin{bmatrix}
1 & 0 \\
0 & 1 \\
1 & 1 \\
1 & 2 \\
2 & 3
\end{bmatrix}, \qquad
\bfH_S =
\begin{bmatrix}
-1 & -1 & 1 & 0 & 0 \\
-1 & -2 & 0 & 1 & 0 \\
-2 & -3 & 0 & 0 & 1 \\
\end{bmatrix}.
\end{eqnarray}
where the same precoding matrix is applied to the noise symbols in each query row. For example, in the first query row, $r_1, r_2$ is precoded by ${\bf C}_S$ to produce $r_1, r_2, r_1 + r_2, r_1 + 2r_2, 2r_1 + 3r_2$ (one for each server). 2-security follows by noting that ${\bf C}_S$ is an MDS matrix where any 2 rows are linearly independent.

In what follows, we present the precoding design of 
$a_{(1:15)}$ and $b_{(1:15)}$. The user privately chooses 2 matrices $\bfU_1$ and $\bfU_2$ uniformly and independently from {\color{black} $GL_9(\Fq)$.} 
The precoding of the desired message symbols occurs in each query row, by the same $5 \times 3$ matrix $\bfG_1$, {
\begin{eqnarray}
\bfG_1 = \begin{bmatrix}
1 & 0 & 0 \\
0 & 1 & 0 \\
0 & 0 & 1 \\
1 & 3 & 2 \\
2 & 3 & 1
\end{bmatrix}. \label{eqn:ex4G1}
\end{eqnarray}
}
The precoding of the undesired message symbols is over all 15 $b_i$ symbols, the coefficients of which are described by a $15 \times 9$ matrix $\bfG_2$. Thus, $a_{(1:15)}$ and $b_{(1:15)}$ are generated by
\begin{eqnarray}
[a_1, a_4, a_7, a_{10}, a_{13}]^T & = & \bfG_1  \bfU_1 [(1:3),:] W_1, \label{eqn:ex4W1precode1} \\
{ [a_2, a_5, a_8, a_{11}, a_{14}]^T } & = & \bfG_1  \bfU_1 [(4:6),:] W_1, \label{eqn:ex4W1precode2} \\
{ [a_3, a_6, a_9, a_{12}, a_{15}]^T } & = & \bfG_1  \bfU_1 [(7:9),:] W_1, \label{eqn:ex4W1precode3} \\
b_{(1:15)} & = & \bfG_2 \bfU_2 W_2. \label{eqn:ex4W2precode}
\end{eqnarray}
The matrix ${\bf G}_2$ will be specified later.

Upon receiving the answers, the user first cancels the noise by multiplying 
$\bfH_S$ with the answers in each query row. 
\begin{center}
\begin{tabular}{|c|c|c|}
\hline
 \small{Server 3 $-$ Server 1 $-$ Server 2} & \small{Server 4 $-$ Server 1 $-$ 2Server 2} & \small{Server 5 $-$ 2Server 1$-$ 3Server 2} \\
\hline
 $a_7 - a_1 - a_4$  &       $a_{10} -a_1 -2a_4$ & $a_{13} -2a_1-3a_4$ \\
$b_7 - b_1 - b_4$  &       $b_{10} -b_1-2b_4$  & $b_{13}-2b_1-3b_4 $ \\
 $a_{8}- a_2-a_5 + b_{8}-b_2-b_5$    & $a_{11} -a_2-2a_5+ b_{11}-b_2-2b_5$ & $a_{14}-2a_2-3a_5 + b_{14}-2b_2-3b_5$ \\
$a_{9}-a_3-a_6 + b_{9}-b_3-b_6$    & $a_{12}-a_3-2a_6 + b_{12}-b_3-2b_6$  & $a_{15}-2a_3-3a_6 + b_{15}-2b_3-3b_6$ \\
\hline
\end{tabular}
\end{center}
where after the noise symbols are cancelled, the $9 \times 9$ precoding matrices associated with $\bfU_1 W_1$ and $\bfU_2 W_2$ respectively are\footnote{From~\eqref{eqn:ex4W1precode1}-\eqref{eqn:ex4W2precode}, $\bfF_1$ is associated with $\bfU_1 W_1$ in an order specified by the query rows (e.g., the coefficients of the first two queries $a_7 - a_1 - a_4$ and $a_{10} -a_1 -2a_4$ in the first query row are specified by the first two rows of $\bfF_1$, respectively); $\bfF_2$ is associated with $\bfU_2 W_2$ in lexicographic order (e.g., the coefficients of the two queries with smallest symbol indices in $b$, $b_7 - b_1 - b_4$ and $b_{8}-b_2-b_5$ are specified by the first two rows of $\bfF_2$, respectively). Note that these orders are not essential in the later proof.}
\begin{eqnarray}
\bfF_1 = (\bfH_S \cdot \bfG_1) \otimes \bfI^3,  \label{eqn:ex4F1}\\
\bfF_2 = (\bfH_S \otimes \bfI^3) \cdot \bfG_2. \label{eqn:ex4F2}
\end{eqnarray}
To successfully decode the 9 symbols of $W_1$, the $b_i$ symbols in the third and fourth query rows should be cancelled by the symbols in the second query row; the 9 $a_i$ symbols should be linearly independent. Therefore, the correctness criterion on $\bfF_1$ and $\bfF_2$ is
\begin{quote}
 {\bf Correctness condition 1 (for desired symbols):} $\bfF_1$ is non-singular.
 
 {\bf Correctness condition 2 (for undesired symbols):} there exists a systematic MDS matrix 
$\bfM_{\text{interference}} = 
\begin{bmatrix}
\bfI^{3} \\
\bfM_1 \\
\bfM_2
\end{bmatrix}
$ where $\bfM_1$ and $\bfM_2$ are 2 $3 \times 3$ matrices, such that 
\begin{eqnarray}
\bfF_2 = \bfM_{\text{interference}} \cdot \bfF_2[(1:3),:]. \label{eqn:ex4correct2}
\end{eqnarray}
Note that this condition is sufficient but not necessary to cancel the interference. 
\end{quote}

From~\eqref{eqn:ex4correct2}, $\bfG_2$ can be represented in terms of $\bfM_1$ and $\bfM_2$,
\begin{eqnarray}
\bfG_2 = 
\begin{bmatrix}
\bfI^9 \\
\bfI^3 - \bfM_1, 2\bfI^3 - \bfM_1, \bfM_1 \\
2\bfI^3 - \bfM_2, 3\bfI^3 - \bfM_2, \bfM_2 
\end{bmatrix} \cdot \bfG_2[(1:9), :],  \label{eqn:ex4G2M1M2}
\end{eqnarray}
where we set $ \bfG_2[(1:9), :]  = {\bf I}^9$.

To guarantee $3$-privacy, we require every 3 servers observe independent random linear combinations of both messages, which is translated into the following two conditions.
\begin{quote}
 {\bf Privacy condition 1 (for desired symbols):} $10$ submatrices of $\bfG_1$, each of which is comprised of 3 distinct rows, are non-singular. 
 
 {\bf Privacy condition 2 (for undesired symbols):} 10 submatrices of $\bfG_2$, each of which is comprised of 9 rows that specify the precoding coefficients of the $b_i$ symbols observed by 3 distinct servers, are non-singular. 
\end{quote}

{\color{black} It is easy to verify that the 10 submatrices of $\bfG_1$ (refer to~\eqref{eqn:ex4G1}) are invertible, and privacy condition 1 is satisfied. Further, \begin{eqnarray}
\bfH_S \cdot \bfG_1 = \begin{bmatrix}
-1 & -1 & 1 \\
0 & 1 & 2 \\
1 & 0 & 1 
\end{bmatrix},
\end{eqnarray}
which is invertible. Hence, $\bfF_1 = (\bfH_S \cdot \bfG_1) \otimes \bfI^3$ is invertible, and correctness condition 1 is satisfied.}

Instead of finding an explicit choice of $\bfG_2$ as in previous sections (which is complicated in general {\color{black}because $\bfG_2$ contains certain non-trivial structure, e.g., see \eqref{eqn:ex4G2M1M2}}), we show the existence of such $\bfG_2$ over a large field via the Schwartz Zippel lemma~\cite{schwartz1980fast,zippel1979probabilistic}. In fact, by choosing the entries of $ {\bf M}_1, {\bf M}_2$ uniformly and independently from a sufficiently large field, the conditions are satisfied with high probability so that almost all choices will work.

Note that $\bfG_2$ is determined by $\bfM_1$ and $\bfM_2$ and we will choose the 18 entries in $\bfM_1$ and $\bfM_2$ independently and uniformly over a sufficiently large field. Correctness condition 2 requires $\bfM_{\text{interference}} = [\bfI^{3} ; \bfM_1 ; \bfM_2]$ to be MDS, meaning that its ${9 \choose 3} $ submatrices (each of which contains 3 distinct rows from $\bfM_{\text{interference}}$) are invertible. It is easy to see that the determinant polynomial for each of them is not the zero-polynomial, then they are invertible with high probability by the Schwartz Zippel lemma~\cite{schwartz1980fast,zippel1979probabilistic}. 
Privacy condition 2 requires 10 specific submatrices of $\bfG_2$ to be invertible. Similarly, we need to show that their determinant polynomials are non-zero, for which we defer the detailed proof to Section~\ref{sec:achievable_general} for the general case to avoid repetition.

Therefore, we have proved the existence of certain ${\bf G}_1, {\bf G}_2$ that produces a private and correct scheme.

\subsection{Servers observe linearly dependent symbols when $T>N-E$, $K=2, N=4, T=3, E=2$} \label{sec:achieve_ex5}
We consider the smallest case where $T > N-E$. In this regime,  it turns out that the servers observe linearly dependent symbols. This phenomenon has not been observed in the literature and existing privacy proofs will fail. We tackle this challenge by increasing the message size ({\it i.e.,} append dummy variables that are globally known) so that the symbols (after lengthening) become linearly independent again.

Assume each message consists of 4 symbols from a field $\Fq$, 
and the user wants $W_1$. The scheme is described below.
\begin{center}
\begin{tabular}{|c|c|c|c|}
\hline
\small{Server 1} & \small{Server 2} & \small{Server 3} & \small{Server 4}  \\
\hline
$a_1 + r_1$           & $a_3 + r_2$           & $a_5 + r_1+ r_2$  &       $a_7 +  r_1 +2r_2$ \\
$b_1 + s_1$           & $b_3 +s_2$           & $b_5 + s_1+s_2$  &       $b_7 + s_1+2s_2$ \\
$a_2 + b_2 + t_1$     & $a_4 + b_4 + t_2$     & $a_{6} + b_{6}+ t_1+t_2$    & $a_{8} + b_{8}+ t_1+2t_2$ \\
\hline
\end{tabular}
\end{center}
where
the precoding matrix of the noise symbols $r_1, r_2, s_1, s_2, t_1,t_2$ and the parity check matrix are
\begin{eqnarray}
\bfC_S = 
\begin{bmatrix}
1 & 0 \\
0 & 1 \\
1 & 1 \\
1 & 2
\end{bmatrix},  \qquad
\bfH_S =
\begin{bmatrix}
-1 & -1 & 1 & 0 \\
-1 & -2 & 0 & 1
\end{bmatrix}.
\end{eqnarray}

%
%
%

Note that every 3 servers observe 6 (mixed) symbols from each message and each message has 4 symbols, so the colluding servers do observe linear dependency in the symbols. To deal with this, we add 2 dummy variables (known to the user) to each message. 
Denote the extended messages by $\tilde{W}_1$ and $\tilde{W}_2$, each of length 6.

The user privately chooses 2 matrices $\bfU_1$ and $\bfU_2$ uniformly and independently from {\color{black} $GL_6(\Fq)$,} 
%
$a_{(1:8)}$ and $b_{(1:8)}$ are generated by 
{\color{black} 
\begin{eqnarray}
a_{(1:8)} &=& \overline{\bfG}_1 \bfU_1 \tilde{W}_1, \\
b_{(1:8)} &=& \bfG_2 \bfU_2 \tilde{W}_2,
\end{eqnarray} 
where $\overline{\bfG}_1$ is an $8 \times 6 $ matrix where each entry is chosen independently and uniformly from a sufficiently large field,
}
and 
\begin{eqnarray}
\bfG_2^{8\times 6} &=& 
\begin{bmatrix}
\bfI^6 \\
\bfI^2 - \bfM, 2\bfI^2-\bfM, \bfM
\end{bmatrix}, \\
\bfM &=& 
\begin{bmatrix}
1 & 1 \\
\psi_1 & \psi_2
\end{bmatrix}
\cdot \text{diag}(\psi_1^2, \psi_2^2) \cdot
\begin{bmatrix}
1 & 1 \\
\psi_1 & \psi_2
\end{bmatrix}^{-1}, ~~~\psi_1 \neq \psi_2, \psi_i^2 - 1 \neq 0, \psi_i^2 - 2 \neq 0. \label{eq:mcon}
\end{eqnarray}


{\bf $3$-privacy:}
{\color{black} 
Note that the 4 submatrices of $\overline{\bfG}_1$ corresponding to 4 possible 3-colluding servers have non-zero determinant polynomials, then by the Schwartz Zippel lemma, 
the probability that $\overline{\bfG}_1$ works approaches 1 as $q$ increases. 
With any such realization of $\overline{\bfG}_1$,
every 3 servers observe linearly independent combinations of $\tilde{W}_1$.
} We may similarly check that 
every 3 servers observe linearly independent combinations of $\tilde{W}_2$.
Hence, $3$-privacy is guaranteed.

{\bf Correctness:} 
After the noise symbols are cancelled, we have 
\begin{center}
\begin{tabular}{|c|c|}
\hline
\small{ Server 3 $-$ Server 1 $-$ Server 2} & \small{ Server 4 $-$ Server 1 $-$ 2Server 2} \\
\hline
$a_5 -a_1-a_3$  &       $a_7-a_1-2a_3 $ \\
 $b_5-b_1-b_3$  &       $b_7-b_1-2b_3 $ \\
   $a_{6}-a_2-a_4 + b_{6}-b_2-b_4$    & $a_{8}-a_2-2a_4 + b_{8}-b_2-2b_4$ \\
\hline
\end{tabular}
\end{center}
where the coefficients of the $b$ symbols are
\begin{eqnarray}
\bfF_2 = (\bfH_S \otimes \bfI^2) \cdot \bfG_2  = 
\begin{bmatrix}
-\bfI & -\bfI & \bfI \\
-\bfM & -\bfM &\bfM
\end{bmatrix} 
 = 
\begin{bmatrix}
\bfI \\
\bfM
\end{bmatrix} \cdot \bfF_2[(1:2),:],
\end{eqnarray}
where $[\bfI;\bfM]$ is MDS from (\ref{eq:mcon}). Then $b_{6}-b_2-b_4$ and $b_{8}-b_2-2b_4$ can be constructed from  $b_5-b_1-b_3$ and $b_7-b_1-2b_3 $.
{\color{black} The coefficients of the 4 desired symbols of $W_1$ 
are
\begin{eqnarray}
\overline{\bfF}_1 \cdot \bfU_1[:, (1:4)] = (\bfH_S \otimes \bfI^2) \cdot \overline{\bfG}_1 \cdot \bfU_1[:, (1:4)].
\end{eqnarray}
To correctly decode the desired symbols, $\overline{\bfF}_1 \cdot \bfU_1[:, (1:4)]$ must be non-singular. Consider an arbitrary realization of $\bfU_1$. Note that $\bfU_1[:, (1:4)]$ has full-column rank, i.e., there exists 4 rows of $\bfU_1[:, (1:4)]$ that are linearly independent (assume to be the first 4 rows without loss of generality). 
Then, we can set the last 2 columns of $\overline{\bfG}_1$ to be zeros, and $\overline{\bfG}_1[:,(1:4)]$ to $\bfG_1 \otimes \bfI^2$, where $\bfG_1 = \begin{bmatrix}
1 & 0 \\
0 & 1 \\
2 & 2 \\
2 & 4
\end{bmatrix}$.
As a result, $(\bfH_S \otimes \bfI^2) \cdot \overline{\bfG}_1[:,(1:4)]   = (\bfH_S \cdot \bfG_1) \otimes \bfI^2$, where $\bfH_S \cdot \bfG_1 = \begin{bmatrix}
1 & 1 \\
1 & 2
\end{bmatrix}$ is invertible. Therefore, $(\bfH_S \otimes \bfI^2) \cdot \overline{\bfG}_1[:,(1:4)]  $ and $\overline{\bfF}_1 \cdot \bfU_1[:, (1:4)]$ are invertible, and the determinant polynomial of $\overline{\bfF}_1 \cdot \bfU_1[:, (1:4)]$ is not the zero polynomial. By the Schwartz Zippel lemma, the determinant polynomial evaluates to a nonzero value with probability approaching 1 as $q$ increases. Averaging over all possible choices of $\bfU_1$, the probability of $\overline{\bfF}_1 \cdot \bfU_1[:, (1:4)]$ being non-singular (hence we can decode correctly) tends to 1 as $q$ increases. 
}

Security and capacity-achieving are easy to verify. The proof is thus complete.

\subsection{General parameters $K, N, T, E$} \label{sec:achievable_general}
In this section, we present a capacity-achieving scheme for general parameters of $K, N, T$ and $E$ when $E < T$. We first introduce the structure of the retrieval scheme and then discuss the detailed precoding of the downloaded symbols. 

\subsubsection{Structure of the retrieval scheme} \label{sec:scheme_structure}
\begin{table}
\begin{center}
{\renewcommand{\arraystretch}{1.4}
\begin{tabular}{|c|c|}
\hline
 & \small{Server $n$ }  \\
\hline
\multirow{3}{*}{Layer 1:} & $ \big\{  X^{[1]}_{i}   +   S_{i}^{[1]}  \big\}_{i \in \calI(n,1,1)}$   \\
                                       & $\vdots$  \\
                                       & $ \big\{ X^{[K]}_{i}   +   S_{i}^{[K]} \big\}_{i \in \calI(n,1,K)}$  \\
\hline
\multirow{8}{*}{Layer 2:} & $\big\{  X^{[1]}_{i} + X^{[2]}_{i}   +   S_{i}^{[1,2]} \big\}_{i \in \calI(n,2,1)} $  \\
                                       & $\big\{  X^{[1]}_{i} + X^{[3]}_{i}   +   S_{i}^{[1,3]} \big\}_{i \in \calI(n,2,2)} $  \\
                                       & $\vdots$  \\
                                       & $\big\{  X^{[1]}_{i} + X^{[K]}_{i}   +   S_{i}^{[1,K]} \big\}_{i \in \calI(n,2,K-1)}$  \\
                                       & $\big\{  X^{[2]}_{i} + X^{[3]}_{i}   +   S_{i}^{[2,3]} \big\}_{i \in \calI(n,2,K)}$  \\
                                       & $\vdots$  \\
                                       & $\big\{  X^{[K-1]}_{i} + X^{[K]}_{i}   +   S_{i}^{[K-1,K]}  \big\}_{i \in \calI(n,2,{K \choose 2})}$  \\
\hline
Layers $3:K-1$ & $\vdots$ \\
\hline
Layer $K$        & $\big\{   X^{[1]}_{i} + X^{[2]}_{i} + \dots + X^{[K]}_{i}   +   S_{i}^{[1:K]} \big\}_{i \in \calI(n,K,1)}$  \\
\hline
\end{tabular} 
}
\caption{The structure of the downloaded symbol from Server $n$.}
 \label{table:generalschemeDBn}
\end{center}
\end{table}

\begin{table}
\begin{center}
{\renewcommand{\arraystretch}{1.4}
\resizebox{18.0cm}{!}{
\begin{tabular}{|c|c|c|c|}
\hline
Type index & Type of $\layer$-sum & Index set & Downloaded symbols \\
\hline
1 & $\{1,2, \dots , \layer -2, \layer -1, \layer  \}$ & $\calI(n, \layer, 1) $ & $\big\{   X^{[1]}_{i} + X^{[2]}_{i} + \dots + X^{[\layer]}_{i}   +   S_{i}^{[1:\layer]} \big\}_{i \in \calI(n,\layer ,1)}$ \\
\hline
2 & $\{1,2, \dots , \layer -2, \layer -1, \layer+1  \}$ & $\calI(n, \layer, 2) $ & $\big\{   X^{[1]}_{i} + \dots +X^{[\layer-1]}_{i}+ X^{[\layer+1]}_{i}   +   S_{i}^{[1,2,\dots,\layer-1, \layer+1]} \big\}_{i \in \calI(n,\layer ,2)}$ \\
\hline
$\vdots$ & $\vdots$ & $\vdots$ &$\vdots$ \\
\hline
$t$ & $\{ t_1, t_2, \dots, t_{\layer} \}$ & $\calI(n, \layer, t) $ & $\big\{   X^{[t_1]}_{i} + X^{[t_2]}_{i} +\dots + X^{[t_{\layer}]}_{i}   +   S_{i}^{[t_1, t_2, \dots, t_{\layer}]} \big\}_{i \in \calI(n,\layer ,t)}$ \\
\hline
$\vdots$ & $\vdots$ & $\vdots$ &$\vdots$ \\
\hline
${K \choose \layer}$ & $\{ K-\layer +1, K-\layer +2, \dots, K \}$ &  $\calI(n, \layer, {K \choose \layer}) $ & $\big\{   X^{[K-\layer +1]}_{i} + X^{[K-\layer +2]}_{i} +\dots + X^{[K]}_{i}   +   S_{i}^{[K-\layer +1:K]} \big\}_{i \in \calI(n,\layer ,{K \choose \layer})}$ \\
\hline
\end{tabular}
}
}
\caption{The structure of layer $\layer$ from Server $n$. $|\calI(n, \layer, t)| = (N-T)^{\layer -1}(T-E)^{K-\layer}$.}
 \label{table:generalschemeLayer}
\end{center}
\end{table}

In our scheme, we force symmetry across the servers, {\it i.e.,} the structure of the downloaded symbols from all servers is the same. Table~\ref{table:generalschemeDBn} shows the detailed structure for Server $n$ and the structure is explained in detail as follows.

The downloaded symbols from Server $n$ may be divided into $K$ layers. For layer $\layer \in [1:K]$, each symbol is a mixture of symbols from $\layer$ distinct messages (hence called an {\bf \emph{$\layer$-sum}}) and a distinct noise symbol. In particular, $X^{[k]}_{i}$ denotes a mixture of symbols from message $W_k$ (corresponding to $a_i$ for a mixture of symbols from $W_1$ in the examples), and $S^{[\calK]}_{i}$ denotes a noise symbol added to an {$\layer$-sum} with type $\calK \subset [1:K]$ (where the type $\calK$ is the set of message indices in the {{$\layer$-sum}}). 
Each layer $\layer$ includes all ${K \choose \layer}$ different types, and for each type, a same number of {$\layer$-sums} are downloaded.
The types are ordered lexicographically and assigned a type index $t$ (see Table~\ref{table:generalschemeLayer}). 
For each type $\{ t_1, t_2, \dots, t_{\layer} \} \subset [1:K]$ with type index $t$, the user downloads $|\calI(n, \layer, t) |$ $\layer$-sums $\{ X^{[t_1]}_{i} + X^{[t_2]}_{i} +\dots + X^{[t_{\layer}]}_{i}   +   S_{i}^{[t_1, t_2, \dots, t_{\layer}]} \}$, where $i \in \calI(n,\layer ,t)$,
\begin{eqnarray}
|  \calI(n, \layer, t)  | \triangleq  I_{\layer} =  (N-T)^{\layer - 1} (T-E)^{K-\layer},
\end{eqnarray}
and $ \calI(n, \layer, t) $ denotes the index set of the {$\layer$-sums}. All the indices are formulated by 
\begin{eqnarray}
\{ \calI(n, \layer, t) \}_{n = [1:N], \layer = [1:K], t = [1: {K \choose \layer} ]} = [1: N L_n] ,
\end{eqnarray}
where $L_n$ is calculated below in~\eqref{eqn:Ln} and $N L_n$ is the total number of mixtures of any message appeared in all $N$ servers. The indices are enumerated from 1 to $N L_n$, and are distributed from Server 1 to Server $N$, and at each server from layer $1$ to layer $K$. (The same indices assignment rule is used in the examples.)



We calculate the rate achieved.
The number of symbols downloaded from each server is
\begin{eqnarray}
D_n = \sum_{\layer = 1}^{K}  {K \choose \layer } \cdot  I_{\layer} = \sum_{\layer = 1}^{K}  {K \choose \layer } \cdot (N-T)^{\layer - 1} (T-E)^{K-\layer} = \frac{(N-E)^K-(T-E)^K}{N-T}.
\end{eqnarray}

For any message $W_k$, the number of mixtures $X^{[k]}_i$ of symbols from $W_k$ involved in the downloads from each server is
\begin{eqnarray}
L_n = \sum_{\layer = 1}^{K}  {K-1 \choose \layer-1 } \cdot  I_{\layer} = \sum_{\layer = 1}^{K}  {K-1 \choose \layer-1 } \cdot (N-T)^{\layer - 1} (T-E)^{K-\layer} = (N-E)^{K-1}. \label{eqn:Ln}
\end{eqnarray} 

In Section~\ref{sec:achieve_precoding} below, we will show that with proper precoding design of the mixtures $\{ X^{[k]}_i \}$ and the noise symbols $\{ S^{[\calK]}_i \}$, the scheme can decode $(N-E) \cdot L_n = (N-E)^K$ symbols from the desired message. Hence, the scheme achieves the rate 
\begin{eqnarray}
R = \frac{(N-E) \cdot L_n}{N \cdot D_n} = \left( 1-\frac{E}{N} \right) \cdot \frac{(N-E)^{K-1}}{\frac{(N-E)^K-(T-E)^K}{N-T}} = \left( 1-\frac{E}{N} \right) \cdot \frac{1 - \frac{T-E}{N-E}}{ 1 - \left( \frac{T-E}{N-E} \right)^K }.
\end{eqnarray}

In the following, we present the details of the precoding design and the justification of security, privacy, and correctness.

\subsubsection{Precoding design} \label{sec:achieve_precoding}
Suppose each message consists of $L = (N-E)^K$ symbols from a sufficiently large field $\Fq$. 
If $T >N-E$, for ease of presentation, we add $(T+E-N)(N-E)^{K-1}$ dummy variables to each message to extend the message length to $ T(N-E)^{K-1}$. 
We abuse the notation by letting $W_k$ denote the extended message. 
Define $\Neff = \max(N-E,T)$ and $L' = L_n \cdot \Neff$.

{\bf Precoding of the noise symbols:}
From Table~\ref{table:generalschemeDBn} in Section~\ref{sec:scheme_structure}, the scheme downloads $D_n$ symbols from each server and each symbol is  {\color{black} added} with a noise symbol. The noise symbols from each query row are precoded by {\color{black} an $(N,E)$-MDS matrix $\bfC_S$.} 
Then the corresponding systematic parity check matrix is denoted by $\bfH_S $. From the dimension of the MDS code, we know that in total, $E \cdot D_n$ independent noise symbols are used, matching the minimum required size for common randomness. The MDS property of the noise precoding matrix translates to $E$-security.

{\bf Precoding of the desired symbols:}
Suppose the user wants to retrieve $W_l$. The user privately chooses $K$ matrices $\bfU_1, \bfU_2, \dots, \bfU_K$ independently and uniformly from {\color{black} $GL_{L'}(\Fq)$.} 
 Let $\overline{\bfG}_l^{N L_n \times \Neff L_n}$ be the precoding matrix of all $X^{[l]}_{i}$ symbols from $W_l$. Thus, all $N \cdot L_n$ mixtures of symbols from $W_l$ in the retrieval scheme are generated by
\begin{eqnarray}
\{ X^{[l]}_{i} \}_{i \in \calI([1:N],:,:)}= \overline{\bfG}_l^{N L_n \times \Neff L_n} \cdot \bfU_l W_l, \label{eqn:precoding_desired_general}
\end{eqnarray}
and then $\{ X^{[l]}_{i} \}_{i \in \calI([1:N],:,:)}$ are distributed evenly to each query row. 
To guarantee $T$-privacy, we require every $T$ servers observe linearly independent random mixtures of symbols from $W_l$. 
That is
\begin{quote}
{\bf Privacy condition 1 (for desired symbols): } The ${N \choose T}$ submatrices of $\overline{\bfG}_l^{N L_n \times \Neff L_n}$, each of which is comprised of $T L_n$ rows corresponding to a set of $T$ servers, are non-singular.
\end{quote}

From all the received symbols, the user cancels the noise by projecting the $N$ symbols in each query row to the {\color{black} dual} space of $\bfC_S$ ({\it i.e.,} multiplying with $\bfH_S$). The user further cancels the interference of undesired symbols (to be justified in the following). The coefficients of $\bfU_l W_l$ are given by $\overline{\bfF}_l = \left( \bfH_S^{(N-E) \times N} \otimes \bfI^{L_n} \right) \cdot \overline{\bfG}_l^{N L_n \times \Neff L_n}$. Therefore, we have the following condition for the correct decoding of $W_l$,
\begin{quote}
{\bf Correctness condition 1 (for desired symbols):} The $L \times L$ square matrix $\overline{\bfF}_l \cdot \bfU_l[:,(1:L)] = \left( \bfH_S^{(N-E) \times N} \otimes \bfI^{L_n} \right) \cdot \overline{\bfG}_l^{N L_n \times \Neff L_n} \cdot \bfU_l[:,(1:L)]$ is non-singular.
\end{quote}

In the following, we show explicit precoding design for the case $T \leq N-E$, and a random choice will work almost surely $T > N-E$.

{\bf Case 1 (Explicit precoding design when $T \leq N-E$):}
In this case, $\Neff = N-E$. Let the precoding matrix $\bfC_S$ of the noise symbols be the Vandermonde matrix $\bfV^N(\lambda_1, \dots, \lambda_E)$. Further, let each query row be precoded by the same matrix $\bfG_l^{N \times (N-E)}$, where $\bfG_l$ is the Vandermonde matrix $\bfV^N(\phi_1, \dots, \phi_{N-E})$. 
The precoding matrix in~\eqref{eqn:precoding_desired_general} is $\overline{\bfG}_l^{N L_n \times \Neff L_n} =  \bfG_l^{N \times (N-E)} \otimes \bfI^{L_n} $.
Then $\{ X^{[l]}_i \}$ are independent across different query rows. Further, every $T$ rows of $\bfG_l$ are linearly independent. Therefore, every $T$ servers observe linearly independent random mixtures of symbols from $W_l$, and privacy condition 1 is satisfied. 

In this case, $L' = L$, and $\bfU_l[:,(1:L)] = \bfU_l$ is non-singular. Note that $\overline{\bfF}_l = \left( \bfH_S^{(N-E) \times N} \otimes \bfI^{L_n} \right) \cdot \left( \bfG_l^{N \times (N-E)} \otimes \bfI^{L_n} \right) = (\bfH_S \cdot \bfG_l) \otimes \bfI^{L_n}$, so correctness condition 1 translates to the requirement of $\bfH_S \cdot \bfG_l$ being non-singular. This ensures that for any realization of ${\bfU}_l$, the user obtains $N-E$ linearly independent combinations in desired symbols from each query row. 
We require that $\{ \lambda_1, \dots, \lambda_E, \phi_1, \dots, \phi_{N-E} \}$ are distinct nonzero elements from $\F_q$ such that the columns of $\bfC_S$ and $\bfG_l$ are linearly independent. 
Note that the parity check matrix $\bfH_S$ projects any vector that is in the span of the columns of $\bfC_S$ to zero. As the columns of $\bfC_S$ and $\bfG_l$ are linearly independent, we know that $\bfH_S \cdot \bfG_l$ is non-singular, and correctness condition 1 is satisfied.

{\color{black}
{\bf Case 2 (Random choices work almost surely when $T > N-E$):}
In this case, $\Neff = T$. Let each entry of $\overline{\bfG}_l^{N L_n \times T L_n}$ be chosen independently and uniformly over $\Fq$. We show that as $q \to \infty$, the probability that $\overline{\bfG}_l$ satisfies correctness condition 1 approaches 1.

The proof relies on the Schwartz Zippel lemma~\cite{schwartz1980fast,zippel1979probabilistic} about the roots of polynomials, where the variables of the polynomials to be considered are the entries of $\overline{\bfG}_l$. 

Privacy condition 1 requires that ${N \choose T}$ submatrices of $\overline{\bfG}_l$ corresponding to ${N \choose T}$ possible sets of $T$-colluding servers are non-singular. It is evident that the determinant polynomial of each submatrix is not the zero polynomial. Therefore, as the field size $q$ increases, the probability that each determinant polynomial evaluates to a nonzero value (the submatrix being non-singular) approaches 1. It follows that all ${N \choose T}$ submatrices are non-singular with probability approaching 1 as $q$ increases. As a uniform choice of $\overline{\bfG}_l$ works almost surely, there are infinite many choices of feasible $\overline{\bfG}_l$. The final choice of $\overline{\bfG}_l$ will be uniformly chosen from this feasible set.

To check correctness condition 1, consider an arbitrary realization of $\bfU_l$. As $\bfU_l$ is a full-rank matrix, there exist $L$ linearly independent rows in $\bfU_l[;,(1:L)]$, the indices of which are denoted by $\calL_{\bfU_l}$. We argue that the determinant polynomial of $\overline{\bfF}_l \cdot \bfU_l[:,(1:L)] = \left( \bfH_S \otimes \bfI^{L_n} \right) \cdot \overline{\bfG}_l \cdot \bfU_l[:,(1:L)]$ is not the zero polynomial, because one can set the $L$ columns of $\overline{\bfG}_l $ with indices $\calL_{\bfU_l}$ to be independent of the null space of $\bfH_S$ (similar choice as in Case 1 suffices),  and other columns to be zeros, such that $\overline{\bfF}_l \cdot \bfU_l[:,(1:L)]$ is non-singular. Therefore, by Schwartz Zippel lemma, the determinant polynomial of $\overline{\bfF}_l \cdot \bfU_l[:,(1:L)]$ evaluates to a nonzero value with probability approaching 1 as the field size $q$ increases.
Averaging over all possible choice of $\bfU_l$, the probability of $\overline{\bfF}_l \cdot \bfU_l[:,(1:L)]$ being non-singular (hence we can decode correctly) tends to 1 as $q$ increases.
}

{\bf Precoding of undesired symbols:}
%
We now consider the precoding for the undesired messages, $W_k, k \neq l$.
We wish to ensure that the undesired symbols that appear in an $\layer$-sum (thus from layer $\layer \in [1:K-1]$) with a type that does not contain the desired message $W_l$ are sufficient to reconstruct the undesired symbols that appear in an $\layer+1$-sum (thus from layer $\layer+1$) with a type that contains the desired message $W_l$. If this is satisfied, then all interferences can be cancelled.
Next, we proceed to consider the precoding of these undesired symbols in the pure undesired $\layer$-sums and the $\layer+1$-sums that contain $W_l$. The design for different $\layer$ is done separately and from $\layer = 1$ to $\layer = K-1$. Consider an arbitrary fixed $\layer$. Further, the design for different types of the undesired $\layer$-sum is done separately and in increasing order of the type index. Consider an arbitrary undesired type $\calK = \{ k_1, \dots, k_{\layer }\}$  where $l \notin \calK$.
Assume the lexicographic index of the type $\calK$ among all types with cardinality $\layer$ ({\it i.e.,} in layer $\layer$) is $t$, and the lexicographic index of type $\calK \cup l$ in layer $\layer +1$ is $t'$. To simplify the notation, $X^{[k_1]}_i + \cdots + X^{[k_{\layer}]}_i$ is denoted as $X^{[\calK]}_i$. 
Therefore, we are considering the design of the following symbols.
\begin{center}
{\renewcommand{\arraystretch}{1.5}
\begin{tabular}{|c|c|c|c|}
\hline
  & \small{Server 1} & $\dots$ & \small{Server $N$} \\
\hline
Layer $\layer$:  &  $ \big\{ X^{[\calK]}_{i}+S^{[\calK]}_{i} \big\}_{i \in \calI(1,\layer, t)}$  & $\dots$ & $ \big\{  X^{[\calK]}_{i}+S^{[\calK]}_{i}  \big\}_{i \in \calI(N,\layer, t)}$ \\
\hline
Layer $\layer +1$:  & $\big\{  X^{[l]}_{i}+X^{[\calK]}_{i}+S^{[\calK,l]}_{i} \big\}_{i \in \calI(1,\layer+1, t')}$  & $\dots$ & $\big\{  X^{[l]}_{i}+X^{[\calK]}_{i}+S^{[\calK,l]}_{i} \big\}_{i \in \calI(N,\layer+1, t')} $ \\
\hline
\end{tabular}
}
\end{center}

Recall that 
$|\calI(n,\layer, t)| = I_{\layer} = (N-T)^{\layer - 1} (T-E)^{K-\layer}$, $|\calI(n,\layer, t')| = I_{\layer+1} = (N-T)^{\layer } (T-E)^{K-\layer-1}$, and $\bfU_1, \bfU_2, \dots, \bfU_K$ are uniform
$L' \times L'$ full-rank matrices, where $L' = \Neff \cdot (N-E)^{K-1} $. 
Let $\bfG_{\calK,\layer}$ denote an $N(I_{\layer}+I_{\layer+1}) \times T(I_{\layer}+I_{\layer+1})$ precoding matrix. 
The symbols $\{ X^{[\calK]}_i \}$ in the above table 
are generated by,
\begin{eqnarray}
\{ X^{[\calK]}_{i} \}_{i \in \calI([1:N], [\layer : \layer +1], \{t,t'\} )}= \bfG_{\calK,\layer} \cdot \sum_{k \in \calK} \bfU_k[\calJ_k,
:] W_k,
\end{eqnarray}
where $\calJ_k$ is a vector with $T(I_{\layer}+I_{\layer+1})$ distinct elements in $[1:L']$ that have not been used (by default, the rows of $\bfU_k$ are used from the first to the last). Then $\{ X^{[\calK]}_i \}$ are distributed evenly to the servers.
Note that the precoding matrix $\bfG_{\calK,\layer}$ does not depend on the message index or the type index. In other words, the same precoding matrix is used for all undesired messages and all types. This is important to align the interference so that the interfering sum could be cancelled without knowing the individual terms in the sum.

The dimension of $\bfG_{\calK,\layer}$ is chosen to ensure $T$-privacy, which requires that
each $T$ servers observe linearly independent random combinations of symbols from $W_k$. This translates to the following condition,
\begin{quote}
{\bf Privacy condition 2 (for undesired symbols): } The ${N \choose T}$ submatrices of $\bfG_{\calK,\layer}$, each of which is comprised of $T$ distinct block rows, are non-singular.
\end{quote}
Note that $\bfG_{\calK,\layer}$ can be viewed as a block matrix consisting of $NT$ square submatrices of dimension $(I_{\layer}+I_{\layer+1})$ and $\bfG_{\calK,\layer}^{I_{\layer}+I_{\layer+1}}[i,:]$ is a \emph{block row} that is comprised of a row of square submatrices.


%
%

After multiplying the symbols in the above table with $\bfH_S = \left[   (-\bfP)^{(N-E) \times E} | \bfI^{N-E}  \right]$ to cancel the noise, we have the precoding matrix of $\bfU_k[\calJ_k, 
:] W_k$ as
%
\begin{eqnarray}
\bfF_{\calK,\layer} &=& \left( \bfH_S \otimes \bfI^{I_{\layer}+I_{\layer+1}} \right) \cdot \bfG_{\calK,\layer} \\
&=& \begin{bmatrix}
\bfG_{\calK,\layer}^{I_{\layer}+I_{\layer+1}}[E+1,:] - \bfP[1,:] \cdot \bfG_{\calK,\layer}^{I_{\layer}+I_{\layer+1}}[(1:E),:] \\
\vdots \\
\bfG_{\calK,\layer}^{I_{\layer}+I_{\layer+1}}[N,:] - \bfP[N-E,:] \cdot \bfG_{\calK,\layer}^{I_{\layer}+I_{\layer+1}}[(1:E),:]
\end{bmatrix}.
\end{eqnarray}

To guarantee correct decoding of the desired message symbols, we require that $\bfF_{\calK,\layer}$ has dependent rows in a way such that the interference of $W_k$ can be cancelled.
Specifically, after noise cancellation, there are $(N-E)I_{\layer}$ $\{ X^{[\calK]}_i \}$ symbols in layer $\layer$ and we wish to ensure that they can reconstruct the $(N-E)I_{\layer+1}$ $\{ X^{[\calK]}_i \}$ symbols (mixed with $\{ X^{[l]}_i \}$) in layer $\layer+1$. To this end, we require that the $\{ X^{[\calK]}_i \}$ symbols in layer $\layer$ and layer $\layer+1$ form 
an MDS code. Therefore, we have the following (sufficient but not necessary) condition on correctness decoding,
\begin{quote}
{\bf Correctness condition 2 (for undesired symbols):} There exists an MDS matrix $\Minter $ of dimension $(N-E)(I_{\layer}+I_{\layer+1}) \times (T-E)(I_{\layer}+I_{\layer+1})$, such that 
\begin{eqnarray}
\bfF_{\calK,\layer}  = \Minter \bfF_{\calK,\layer}^{I_{\layer}+I_{\layer+1}} [(1:T-E),:]. \label{eqn:correct2}
\end{eqnarray}
\end{quote} 
Note that $(N-E) \cdot I_{\layer} = (T-E) \cdot (I_{\layer}+I_{\layer+1})$.

In the following, we expand $\Minter$ and~\eqref{eqn:correct2} 
to represent $\bfG_{\calK,\layer}$ by $\Minter$, such that the privacy condition 2 can also be translated into a condition on $\Minter$.

Let $\{ \bfM_{i,j} \}_{i \in [1:N-T], j \in [1:T-E]}$ be $(N-T)(T-E)$ square matrices of dimension $I_{\layer}+I_{\layer+1}$, and $\Minter$ be a systematic MDS matrix such that 
\begin{eqnarray}
\bfM_{\textrm{interference}} = 
\begin{bmatrix}
 & \bigI^{(T-E)(I_{\layer}+I_{\layer+1})} &  \\
\bfM_{1, 1} & \dots & \bfM_{1,T-E} \\
\vdots & \vdots & \vdots \\
\bfM_{N-T, 1} & \dots & \bfM_{N-T,T-E}
\end{bmatrix}. \label{eqn:Minterference}
\end{eqnarray}
Then~\eqref{eqn:correct2} is equivalent to
\begin{align}
& 
\begin{bmatrix}
\bfG_{\calK,\layer}^{I_{\layer}+I_{\layer+1}}[T+1,:] - \bfP[T-E+1,:] \cdot \bfG_{\calK,\layer}^{I_{\layer}+I_{\layer+1}}[(1:E),:] \\
\vdots \\
\bfG_{\calK,\layer}^{I_{\layer}+I_{\layer+1}}[N,:] - \bfP[N-E,:] \cdot \bfG_{\calK,\layer}^{I_{\layer}+I_{\layer+1}}[(1:E),:]
\end{bmatrix} =  \\
&
\begin{bmatrix}
\bfM_{1, 1} & \dots &  \bfM_{1,T-E} \\
\vdots & \vdots & \vdots \\
\bfM_{N-T, 1} & \dots & \bfM_{N-T,T-E}
\end{bmatrix} \cdot
\begin{bmatrix}
\bfG_{\calK,\layer}^{I_{\layer}+I_{\layer+1}}[E+1,:] - \bfP[1,:] \cdot \bfG_{\calK,\layer}^{I_{\layer}+I_{\layer+1}}[(1:E),:] \\
\vdots \\
\bfG_{\calK,\layer}^{I_{\layer}+I_{\layer+1}}[T,:] - \bfP[T-E,:] \cdot \bfG_{\calK,\layer}^{I_{\layer}+I_{\layer+1}}[(1:E),:].
\end{bmatrix}.
\end{align}

This imposes the structure of $\bfG_{\calK,\layer}$ to be
\begin{equation}
\bfG_{\calK,\layer} = 
\underbrace{
\left[
\begin{array}{ccc;{2pt/2pt}c cc}
 \multicolumn{6}{c}{\bigI^{T(I_{\layer}+I_{\layer+1})} } \\
 \hdashline
\bfN_{1,1} & \dots & \bfN_{1,E} & \bfM_{1, 1} & \dots &  \bfM_{1,T-E} \\
  & \vdots &  &  & \vdots &  \\
 \bfN_{N-T,1} & \dots & \bfN_{N-T,E} & \bfM_{N-T, 1} & \dots &  \bfM_{N-T,T-E}
\end{array} 
\right]
}_{\textstyle  \triangleq \tilde{\bfG}_{\calK,\layer}  \mathstrut} 
\cdot \bfG_{\calK,\layer}^{I_{\layer}+I_{\layer+1}}[(1:T),:] , \label{eqn:newGk}
\end{equation}
where $\bfN_{i,j} = 
\bfP[T-E+i,j]\cdot \bfI^{I_{\layer}+I_{\layer+1}} - \sum_{t = 1}^{T-E} \bfP[t,j] \cdot \bfM_{i,t}$. 
We set 
$\bfG_{\calK,\layer}^{I_{\layer}+I_{\layer+1}}[(1:T),:]$ 
as the identity matrix such that $ \bfG_{\calK,\layer} =  \tilde{\bfG}_{\calK,\layer}$.



Therefore, the privacy condition 2 is equivalent to
\begin{quote}
{\bf Privacy condition 2' (for undesired symbols):}  There exist $(N-T)(T-E)$ square matrices $\{\bfM_{i,j}\}_{i \in [1:N-T], j \in [1:T-E]}$ of dimension $I_{\layer}+I_{\layer+1}$, such that 
${N \choose T}$ submatrices of  $\tilde{\bfG}_{\calK,\layer}$, each of which is comprised of distinct $T$ block rows, are non-singular.
\end{quote}

Now correctness condition 2 and privacy condition 2' are both stated in terms of $\bfM_{i,j}$. 
We next proceed to show the existence of $\bfM_{i,j}$ such that both conditions are satisfied. To this end, we will pick each entry of $\bfM_{i,j}$ independently and uniformly from a sufficiently large field.

First, consider correctness condition 2, which requires $\bfM_{\textrm{interference}}$ to be MDS. From (\ref{eqn:Minterference}), it is easy to see that over a sufficiently large finite field, by randomly choosing $\bfM_{i,j}$, $\bfM_{\textrm{interference}}$ is MDS with high probability. 

Second, consider privacy condition 2', which requires the submatrices of $\tilde{\bfG}_{\calK,\layer}$ (made up of $T$ block rows) to be non-singular.
For these submatrices, 
we have three cases. 
\begin{itemize}
\item Case 1: The submatrix consists of the first $T$ block rows, which is an identity matrix and therefore is non-singular directly.

\item Case 2: The submatrix consists of $t$ block rows from the first $T$ block rows, and $T-t$ block rows from the remaining $N-T$ block rows. 
We show that by elementary row and column operations and a proper choice of a realization of $\bfM_{i,j}$, the matrix is non-singular. 

First consider the first $t$ block rows. Each one of the first $t$ block rows contains one identity matrix, and by exchanging columns, we have the identity matrix on the upper left corner and then the upper right corner is a zero matrix (see~\eqref{eqn:Gkcase2M}). It remains to show that the lower right corner (denoted by $\bfM^{T-t}$ in~\eqref{eqn:Gkcase2}) is non-singular. 

The last $T-t$ block rows depend on the $\{ \bfM_{i,j} \}$ matrices. 
Here we have 3 cases: 1) $\bfM^{T-t}$ contains only $\{\bfM_{i,j} \}$ matrices, 2) $\bfM^{T-t}$ contains both $\{ \bfM_{i,j} \}$ and $\{ \bfN_{i,j} \}$ matrices, and 3) $\bfM^{T-t}$ contains only $\{ \bfN_{i,j} \}$ matrices. In the following, we prove the non-singularity of $\bfM^{T-t}$ for all 3 cases.


\begin{eqnarray}
\left[
\begin{array}{c  c c;{2pt/2pt}c  c  c}
\bfI            & {\bf 0}  & {\bf 0} &  &  &  \\
\mathbf{0} & \ddots & {\bf 0} & & \bigZero^{t \times (T-t)} &  \\
\mathbf{0} & \mathbf{0} & \bfI &  & &  \\
\hdashline
  \cdots & \cdots & \cdots  &    &   \bigM^{T-t}   &
\end{array}
\right] \label{eqn:Gkcase2}
\end{eqnarray}

{\it Case 2.1:} $\bfM^{T-t}$ consists of only $\{ \bfM_{i,j} \}$ matrices (this may happen when $t \geq E$) with distinct $(i,j)$ indices and therefore is non-singular by setting it to identity.



{\it Case 2.2:} $\bfM^{T-t}$ consists of $T-t-j$ block columns of $\{ \bfM_{i,j} \}$ matrices and $j$ block columns of $\{ \bfN_{i,j} \}$ matrices, where $\max(E-t,0) \leq j \leq E$. We rearrange $\{ \bfM_{i,j} \}$ matrices to the left so that $\bfM^{T-t}$ is in the form of~\eqref{eqn:Gkcase2M}. Note that this operation does not change the upper right zero matrix in~\eqref{eqn:Gkcase2}. It is easy to see that the upper left part in~\eqref{eqn:Gkcase2M} is non-singular. The lower left part (which contains $\bfM_{i,j}$ matrices) is set to the zero matrix. The lower right part is the Kronecker product of a square submatrix of $\bfP$ (of dimension $j \leq E$) and the identity matrix (note that the $\{ \bfM_{i,j} \}$ matrices in these $\{ \bfN_{i,j} \}$ matrices in the lower right part are set to $\mathbf{0}$ so that we are left with only $\bfP^j$). Remember that $\bfP$ is MDS, so every square submatrix of $\bfP$ is non-singular. Then, $\bfP^{j} \otimes \bfI $ is also non-singular and $\bfM^{T-t}$ has full rank.

\begin{eqnarray}
\bfM^{T-t} = 
\left[
\begin{array}{c  c  c;{2pt/2pt}c }
\bfM_{i_1,1} & \cdots & \bfM_{i_1,T-t-j} & \\
\vdots &  \ddots & \vdots &  \vdots \\
\bfM_{i_{T-t-j},1} & \cdots & \bfM_{i_{T-t-j},T-t-j} & \\
\hdashline
& \bigZero^{j \times (T-t-j)} &  &    \bfP^{j} \otimes \bfI  
\end{array}
\right] \label{eqn:Gkcase2M}
\end{eqnarray}

{\it Case 2.3:} $\bfM^{T-t}$ consists of only $\{ \bfN_{i,j} \}$ matrices (this may happen when $t \geq T-E$) with distinct $(i,j)$ indices. In this case, we set all the $\{ \bfM_{i,j} \}$ matrices which appear in $\{ \bfN_{i,j} \}$ matrices to be zero matrices. Then 
\begin{eqnarray}
\bfM^{T-t} = \bfP^{T-t} \otimes \bfI, 
\end{eqnarray}
where $T-t \leq E$. Similar to the arguments in Case 2.2, $\bfP^{T-t} $ is non-singular and so is $\bfM^{T-t}$.

\item Case 3: The submatrix consists of $T$ block rows from the last $N-T$ blocks rows of $\tilde{\bf G}_{\calK,\layer}$. 
This case corresponds to Case 2.2 when $t=0$, and the submatrix is the same as that in~\eqref{eqn:Gkcase2M} with $t=0$ and $j=E$. 
In this case, we must have $T \leq N-T$ such that $N \geq 2T > 2E$, and $N-E > E$. Then the $E \times E$ submatrix $\bfP^E$ of $\bfP$ 
is non-singular and so is the submatrix.
\end{itemize}

The analysis above shows that the determinant polynomials of the required submatrices of $\tilde{\bf G}_{\calK,\layer}$ are nonzero. Then, by the Schwartz Zippel lemma~\cite{schwartz1980fast,zippel1979probabilistic}, a uniform random choice of $\bfM_{i,j}$ from a sufficiently large field will ensure that with high probability, the determinants of the submatrices are nonzero. Therefore, the submatrices are non-singular and note that this holds for undesired symbols with all types and all layers. Further, for the symbols designed for each type and each two layers, disjoint row vectors from $\bfU_k$ are multiplied with ${\bf G}_{\calK,\layer}$ and in total $TL_n$ vectors from $\bfU_k$ are used. Then by Lemma 1 in \cite{sun2017colluding}, the precoding coefficients for $W_k, k \in [1:K]$ are identically distributed as ${\bfU}_k[(1:TL_n);:]$. 
As ${\bf U}_k$ are identically distributed for all $k$, the message indices are equally likely and can not be distinguished. In other words, any $T$ servers see linearly independent uniformly random combinations of both the desired message and undesired messages, and $T$-privacy is guaranteed.



To sum up, we have completed the description of the precoding of noise, desired and undesired symbols, and have proved that the scheme is secure, correct, private and the rate achieved is equal to the capacity. The achievability proof is thus complete.



\section{Achievability when $E \geq T$} \label{sec:schemeEgreaterT}
Proving achievability in this case is relatively straightforward and follows from that in \cite{wang2017secure}. We present the proof here for completeness.

Suppose each message is comprised of $N-E$ symbols from $\Fq$ with $q > E$. Let $W_{(1:K)} = (W_{1}^{[1]}, \dots, W_{N-E}^{[1]}, \dots, W_{1}^{[K]}, \dots, W_{N-E}^{[K]})$ represent the symbols of all $K$ messages. 
The user wants to retrieve $W_k = (W_{1}^{[k]}, \dots, W_{N-E}^{[k]})$ privately from the servers.

The queries are generated in the following way. 
The user firstly generates $E$ independent uniformly random vectors $U_1, \dots, U_{E}$ of length $K(N-E)$, and then chooses an $(N, E)$-GRS (generalized Reed-Solomon) code with generating matrix $\bfG_{(N,E)}$.
Let $\Lambda = (\lambda_1,\dots,\lambda_N) \in \Fq^N$ where $\lambda_i \neq \lambda_j$. Let $\Phi = (\phi_1, \dots, \phi_N) \in \Fq^N$ where $\phi_i \neq 0$. The generating matrix $\bfG_{(N,E)}$ is defined by
\begin{eqnarray}
\bfG_{(N,E)} = \bfV^E(\lambda_1,  \dots, \lambda_N) 
\cdot \text{diag}(\Phi)
. \label{eqg}
\end{eqnarray}

Let $e_i^{[k]}$ denote the length-$(K(N-E))$ unit vector where only the $\big( (k-1)(N-E)+i \big)$-th entry is $1$ and all the other entries are equal to $0$. The purpose of $e_i^{[k]}$ is to retrieve the $i$-th entry of $W_k$. 
The query vectors are generated by
\begin{eqnarray} 
[Q_1^{[k]}, \dots, Q_N^{[k]}]  = [U_1, \dots, U_{E}] \cdot \bfG_{(N,E)} + [0, \dots, 0, e_1^{[k]}, \dots, e_{N-E}^{[k]}]. \label{eqn:queries_E}
\end{eqnarray}

The nodes share $E$  common randomness symbols $(S_1, \dots, S_{E}) = S_{(1:E)}$ that are uniformly and independently chosen from $\Fq$. 
The servers generate their answers by taking the inner product of the query vector and the stored data vector, then adding a linear combination of the common randomness,
\begin{eqnarray} \label{eqn:answern_E}
A_n^{[k]} = \langle Q_n^{[k]}, W_{(1:K)} \rangle + \langle \bfG_{(N,E)} [:,n], S_{(1:E)} \rangle .
\end{eqnarray}
Let $X_j = \langle U_j, W_{(1:K)} \rangle +S_j, j \in [1:E]$, the answers received by the user can be presented by
\begin{eqnarray} 
[A_1^{[k]}, \dots  , A_N^{[k]}] & =  [X_1,  \dots , X_{E}, W_1^{[k]}, \dots  , W_{N-E}^{[k]}]  \cdot 
\begin{bmatrix}
  \bfG_{(N,E)} \\
\mathbf{0} \; \; \; \; \; \bfI
\end{bmatrix}, \label{eqn:answers_E}
\end{eqnarray}
where the dimensions of  $\mathbf{0}$ and $\bfI$ are clear from the context. 

\noindent {\bf Correctness:} From (\ref{eqg}), $\begin{bmatrix}
  \bfG_{(N,E)} \\
\mathbf{0} \; \; \; \; \; \bfI
\end{bmatrix}$
is invertible such that the user can 
obtain $W_k$.

\noindent {\bf $T$-privacy:} The $T$-privacy constraint is guaranteed, because from~\eqref{eqn:queries_E}, every $E$ query vectors are independently and uniformly distributed. Note that $E \geq T$, so every $T$ nodes can not tell which message the user requests. Note also that $E$-privacy is obtained for free.

\noindent {\bf $E$-security:} 
Guaranteed by the MDS property of the GRS code, {\it i.e.,} any $E$ columns of $\bfG_{(N,E)}$ are linearly independent.

\begin{remark}
Besides the desired message, the user could further obtain $[X_1, \dots, X_{E}]$ from (\ref{eqn:answers_E}). Theses $E$ symbols are protected by independent noise symbols $S_{(1:E)}$. Therefore, the user learns no extra information beyond the desired message, {\it i.e.,} symmetric PIR constraint is satisfied at no additional cost. 
\end{remark}

\section{Converse when $E<T$}
Let us start with a few lemmas.
The lemma below shows that the answers of a set of servers do not depend on the queries when other messages are desired or the queries to other servers, after conditioning on their own queries.

\begin{lemma} \label{thm:noQ}
For any set of nodes $\nodeset \subset [1:N]$, and any set of message index $\calK \subset [1:K]$,
	\begin{eqnarray}
		H(A_{\nodeset}^{[k]} | \queries, W_{\calK}, Q_{\nodeset}^{[k]}) & =  & H(A_{\nodeset}^{[k]} | W_{\calK}, Q_{\nodeset}^{[k]}); \label{eqn:noQ_2} \\
		H(A_{\nodeset}^{[k]} |  W_{\calK}, Q_{[1:N]}^{[k]}) & =  & H(A_{\nodeset}^{[k]} | W_{\calK}, Q_{\nodeset}^{[k]}). \label{eqn:noQ_2new}
	\end{eqnarray}	
\end{lemma}
\noindent {\it Proof:}
%
As $I(A_{\nodeset}^{[k]}; \queries | W_{\calK}, Q_{\nodeset}^{[k]}) \geq  0$ holds trivially, we only need to show
$I(A_{\nodeset}^{[k]}; \queries | W_{\calK}, Q_{\nodeset}^{[k]}) \leq  0$. The proof is presented next. 
\begin{eqnarray}
 I(A_{\nodeset}^{[k]}; \queries | W_{\calK}, Q_{\nodeset}^{[k]}) 
& \leq & I(A_{\nodeset}^{[k]}, W_{[1:K]}, \secrecy ; \queries | W_{\calK}, Q_{\nodeset}^{[k]}) \\
& = & I(W_{[1:K]}, \secrecy ; \queries | W_{\calK}, Q_{\nodeset}^{[k]}) + I(A_{\nodeset}^{[k]} ; \queries |  W_{[1:K]}, \secrecy ,W_{\calK}, Q_{\nodeset}^{[k]}) \\
& \stackrel{(a)}{=} & I(W_{[1:K]}, \secrecy ; \queries | W_{\calK}, Q_{\nodeset}^{[k]}) \\
& = & H(W_{[1:K]}, \secrecy | W_{\calK}, Q_{\nodeset}^{[k]}) - H(W_{[1:K]}, \secrecy| W_{\calK},\queries ) \\
& \stackrel{(b)}{\leq} &  H(W_{[1:K] \setminus {\calK}}, \secrecy ) - H(W_{[1:K] \setminus {\calK}}, \secrecy | \queries) \\ 
& {=}  & I(W_{[1:K] \setminus {\calK}}, \secrecy ; \queries) \\
&\stackrel{(c)}{=}& 0,
\end{eqnarray}
where $(a)$ holds because the answers are deterministic functions of the messages, common randomness, and queries. $(b)$ and $(c)$ follow from the independence of the messages, queries and common randomness (refer to~\eqref{eqn:UQindepW}\eqref{eqn:SindepWQ}).

Note that we allow $\calK$ to be the empty set.~\eqref{eqn:noQ_2new} follows from~\eqref{eqn:noQ_2} and is the form that will be used in the remaining proofs.
\hfill $\Box$

The following lemma allows us to split the answers from all $N$ servers into two parts, one from $E$ servers and the other from the remaining $N-E$ servers.

\begin{lemma} \label{thm:NtoN-E}
For any node set $\calE \subset [1:N]$ and $|\calE| = E$,
\begin{eqnarray}
\left( 1 - \frac{E}{N} \right) H( A_{[1:N]}^{[1]}| Q_{[1:N]}^{[1]} ) \geq L - \smallo(L)
+ H(A_{[1:N] \setminus \calE}^{[1]}| W_1, A_{\calE}^{[1]}, Q_{[1:N]}^{[1]}) .
\end{eqnarray}
\end{lemma}
\noindent {\it Proof:}
From the correctness constraint, $H(W_1 | A_{[1:N]}^{[1]}, Q_{[1:N]}^{[1]}) = \smallo(L)$, we have
\begin{eqnarray}
& \quad & H( A_{[1:N]}^{[1]}| Q_{[1:N]}^{[1]}) \\
& = &H(W_1 | Q_{[1:N]}^{[1]}) + H(A_{[1:N]}^{[1]}| W_1, Q_{[1:N]}^{[1]}) - H(W_1 | A_{[1:N]}^{[1]}, Q_{[1:N]}^{[1]}) \\
& = & L + H(A_{[1:N]}^{[1]}| W_1, Q_{[1:N]}^{[1]}) - \smallo(L) \\
& = & L - \smallo(L)  + H(A_{\calE}^{[1]}| W_1, Q_{[1:N]}^{[1]}) + H(A_{[1:N] \setminus \calE}^{[1]}| W_1, A_{\calE}^{[1]}, Q_{[1:N]}^{[1]}) \\
&  \stackrel{(a)}{=} &  L - \smallo(L)  + H(A_{\calE}^{[1]}| Q_{[1:N]}^{[1]}) + H(A_{[1:N] \setminus \calE}^{[1]}| W_1, A_{\calE}^{[1]}, Q_{[1:N]}^{[1]}) \\
&  \stackrel{(b)}{\geq } &  L - \smallo(L)  + \frac{E}{N} H( A_{[1:N]}^{[1]}| Q_{[1:N]}^{[1]}) + H(A_{[1:N] \setminus \calE}^{[1]}| W_1, A_{\calE}^{[1]}, Q_{[1:N]}^{[1]}),
\end{eqnarray}
where in $(a)$, $\calE$ can be any node set with size $E$. $(b)$ follows by averaging over all $\calE \in [1:N]$ with size $E$, and applying Han's inequality~\cite{cover2012elements}.
\hfill $\Box$

For the answers from $N-E$ servers, we have the following recursive relationship.

\begin{lemma} \label{thm:iteration}
For any $k \in [1:K-1]$, and for any node set $\calE \subset [1:N]$ and $|\calE| = E$,
\begin{eqnarray}
H(A_{[1:N] \setminus \calE}^{[k]} | W_{[1:k]}, A_{\calE}^{[k]}, Q_{[1:N]}^{[k]}) \geq
\frac{T-E}{N-E} \left( H(A_{[1:N] \setminus \calE}^{[k+1]} | W_{[1:k+1]}, A_{\calE}^{[k+1]}, Q_{[1:N]}^{[k+1]})  +L - \smallo(L) \right). \label{eqn:iterationlemma}
\end{eqnarray}
\end{lemma}
\noindent {\it Proof:}
For any set $\calT' \in [1:N] \setminus \calE$ with size $|\calT'|=T-E$,
\begin{eqnarray}
 H(A_{[1:N] \setminus \calE}^{[k]} | W_{[1:k]}, A_{\calE}^{[k]}, Q_{[1:N]}^{[k]}) 
& \geq & H(A_{\calT'}^{[k]} | W_{[1:k]}, A_{\calE}^{[k]}, Q_{[1:N]}^{[k]}) \\
& \stackrel{(a)}{=} & H(A_{\calT'}^{[k]} | W_{[1:k]}, A_{\calE}^{[k]}, Q_{\calT' \cup \calE}^{[k]}) \\
& \stackrel{(b)}{=} & H(A_{\calT'}^{[k+1]} | W_{[1:k]}, A_{\calE}^{[k+1]}, Q_{\calT' \cup \calE}^{[k+1]}) \\
& \stackrel{(c)}{=} &  H(A_{\calT'}^{[k+1]} | W_{[1:k]}, A_{\calE}^{[k+1]}, Q_{[1:N]}^{[k+1]}),
\end{eqnarray}
where $(a)$ and $(c)$ follow from 
Lemma~\ref{thm:noQ}, 
and $(b)$ is due to the privacy constraint~\eqref{eqn:modelprivacy}.

Averaging over all $\calT' \in [1:N] \setminus \calE$ with size $T-E$ and using Han's inequality~\cite{cover2012elements}, we have
\begin{eqnarray}
& \qquad & H(A_{[1:N] \setminus \calE}^{[k]} | W_{[1:k]}, A_{\calE}^{[k]}, Q_{[1:N]}^{[k]})  \\
& \geq & \frac{1}{ {N-E \choose T-E} } \sum_{\substack{ \calT' \subset [1:N] \setminus \calE \\ |\calT'|=T-E}} H(A_{\calT'}^{[k+1]} | W_{[1:k]}, A_{\calE}^{[k+1]}, Q_{[1:N]}^{[k+1]}) \\
& \geq & \frac{T-E}{N-E} \cdot H(A_{[1:N] \setminus \calE}^{[k+1]} | W_{[1:k]}, A_{\calE}^{[k+1]}, Q_{[1:N]}^{[k+1]}) \\
& = & \frac{T-E}{N-E} \cdot \big(   H(A_{[1:N] \setminus \calE}^{[k+1]} | W_{[1:k+1]}, A_{\calE}^{[k+1]}, Q_{[1:N]}^{[k+1]})  + H(W_{k+1} | W_{[1:k]}, A_{\calE}^{[k+1]}, Q_{[1:N]}^{[k+1]}) \\
& \qquad & - H(W_{k+1} | W_{[1:k]}, A_{[1:N]}^{[k+1]}, Q_{[1:N]}^{[k+1]}) \big) \\
& \stackrel{(a)}{=}  &  \frac{T-E}{N-E} \cdot \big(   H(A_{[1:N] \setminus \calE}^{[k+1]} | W_{[1:k+1]}, A_{\calE}^{[k+1]}, Q_{[1:N]}^{[k+1]}) + L - o(L).
\end{eqnarray}
In $(a)$, the second term follows from the 
security constraint~\eqref{eqn:modelsecurity} and the independence of the messages, and the third term follows from 
the correctness constraint~\eqref{eqn:modelcorrect}.

\hfill $\Box$

\subsection{The proof of $R \leq \left( 1- \frac{E}{N} \right)  \left(1 +  \frac{T-E}{N-E} + \cdots + \left( \frac{T-E}{N-E} \right)^{K-1} \right)^{-1}$} \label{sec:converseRcase1}

Applying Lemma~\ref{thm:iteration} iteratively, we have
\begin{align}
& \quad  H(A_{[1:N] \setminus \calE}^{[1]}| W_1, A_{\calE}^{[1]}, Q_{[1:N]}^{[1]}) \\
& \geq  \frac{T-E}{N-E} \big(  H(A_{[1:N] \setminus \calE}^{[2]}| W_1, W_2, A_{\calE}^{[2]}, Q_{[1:N]}^{[2]})  + L - o(L) \big) \\
& \geq  (L - \smallo(L)) \left( \frac{T-E}{N-E} + \left( \frac{T-E}{N-E} \right)^2 \right) + \left( \frac{T-E}{N-E} \right)^2 H(A_{[1:N] \setminus \calE}^{[3]}| W_{[1:3]}, A_{\calE}^{[3]}, Q_{[1:N]}^{[3]}) \\
& \geq  (L - \smallo(L)) \left( \frac{T-E}{N-E} + \cdots + \left( \frac{T-E}{N-E} \right)^{K-1} \right) + \left( \frac{T-E}{N-E} \right)^{K-1} H(A_{[1:N] \setminus \calE}^{[K]}| W_{[1:K]}, A_{\calE}^{[K]}, Q_{[1:N]}^{[K]}) \\
& \geq  (L - \smallo(L)) \left( \frac{T-E}{N-E} + \cdots + \left( \frac{T-E}{N-E} \right)^{K-1} \right) .
\end{align}

Combining with Lemma~\ref{thm:NtoN-E}, we have
\begin{eqnarray}
\left( 1- \frac{E}{N} \right) H(A_{[1:N]}^{[1]}| Q_{[1:N]}^{[1]}) \geq (L - \smallo(L)) \left(1 +  \frac{T-E}{N-E} + \cdots + \left( \frac{T-E}{N-E} \right)^{K-1} \right). \label{eqff}
\end{eqnarray}
From the definition of the PIR rate and by letting $L \to \infty$,
\begin{align}
R = \frac{H(W_1)}{\sum_{n=1}^{N} H(A_n^{[1]} | Q_n^{[1]})} & \leq \frac{L}{H(A_{[1:N]}^{[1]}| Q_{[1:N]}^{[1]})} \\
& \leq \left( 1- \frac{E}{N} \right)  \left(1 +  \frac{T-E}{N-E} + \cdots + \left( \frac{T-E}{N-E} \right)^{K-1} \right)^{-1}.  
\end{align}

\subsection{The proof of $\rho \geq \frac{E}{N-E} \left(1 +  \frac{T-E}{N-E} + \cdots + \left( \frac{T-E}{N-E} \right)^{K-1} \right) $} \label{sec:converserhocase1}
For any set of nodes $\calE \subset [1:N]$ with size $|\calE|=E$, from the security constraint~\eqref{eqn:modelsecurity},
\begin{eqnarray}
0 
& = &  I(W_{[1:K]} ; A_{\calE}^{[k]} , Q_{\calE}^{[k]} ) \\
& =  & H(A_{\calE}^{[k]} | Q_{\calE}^{[k]}) - H(A_{\calE}^{[k]} | W_{[1:K]}, Q_{\calE}^{[k]}) \\
& \stackrel{(a)}{=} & H(A_{\calE}^{[k]} | Q_{\calE}^{[k]})  -  H(A_{\calE}^{[k]} |  W_{[1:K]}, Q_{\calE}^{[k]})  +  H(A_{\calE}^{[k]} | W_{[1:K]}, Q_{\calE}^{[k]}, \secrecy) \\
& =  & H(A_{\calE}^{[k]} | Q_{\calE}^{[k]}) - I(\secrecy ; A_{\calE}^{[k]} | W_{[1:K]}, Q_{\calE}^{[k]}) \\
& \geq & H(A_{\calE}^{[k]} | Q_{\calE}^{[k]}) - H(\secrecy) \\
& \stackrel{(b)}{=} &  H(A_{\calE}^{[k]} | Q_{[1:N]}^{[k]}) - H(\secrecy).
\end{eqnarray}
$(a)$ holds because the answers are deterministic functions of the queries, messages, and common randomness.
By setting $\calK = \emptyset$ in Lemma~\ref{thm:noQ}, we have $(b)$.

Averaging over all $\calE$, 
\begin{eqnarray}
H(\secrecy)
& \geq & \frac{1}{{N \choose E}} \sum_{   \substack{ \calE \in [1:N] \\ |\calE|=E }   } H(A_{\calE}^{[k]} | Q_{[1:N]}^{[k]}) \\
& \geq & \frac{E}{N} \cdot H(A_{[1:N]}^{[k]} | Q_{[1:N]}^{[k]}) . \label{eqn:converserho}
\end{eqnarray}

From (\ref{eqff}), we have 
$H(A_{[1:N]}^{[k]} | Q_{[1:N]}^{[k]}) \geq \frac{N}{N-E} \left(1 +  \frac{T-E}{N-E} + \cdots + \left( \frac{T-E}{N-E} \right)^{K-1} \right) (L - \smallo(L))$.
Hence, by letting $L \to \infty$, $\rho = \frac{H(S)}{H(W_k)} \geq \frac{E}{N-E} \left(1 +  \frac{T-E}{N-E} + \cdots + \left( \frac{T-E}{N-E} \right)^{K-1} \right) $.

\section{Converse when $E \geq T$} \label{sec:conversecase2}
\subsection{The proof of $R \leq 1 - \frac{E}{N}$ } \label{sec:pfThm2_1}
From the correctness constraint, for any message $W_k$, $H(W_k|A_{[1:N]}^{[k]}, Q_{[1:N]}^{[k]}) = \smallo(L)$. For any set of nodes $\calE \subset [1:N]$ with size $|\calE|=E$,
\begin{eqnarray}
H(W_k) 
& = &  H(W_k|Q_{[1:N]}^{[k]}) - H(W_k|A_{[1:N]}^{[k]}, Q_{[1:N]}^{[k]})   + \smallo(L)\\
& = & H(A_{[1:N]}^{[k]} | Q_{[1:N]}^{[k]}) - H(A_{[1:N]}^{[k]} | W_k, Q_{[1:N]}^{[k]}) + \smallo(L)\\
& \leq & H(A_{[1:N]}^{[k]} | Q_{[1:N]}^{[k]}) - H(A_{\calE}^{[k]} | W_k, Q_{[1:N]}^{[k]}) + \smallo(L)\\
& \stackrel{(a)}{=} & H(A_{[1:N]}^{[k]} | Q_{[1:N]}^{[k]}) - H(A_{\calE}^{[k]} | W_k, Q_{\calE}^{[k]}) + \smallo(L)\\
& \stackrel{(b)}{=} & H(A_{[1:N]}^{[k]} | Q_{[1:N]}^{[k]}) - H(A_{\calE}^{[k]} |  Q_{\calE}^{[k]}) + \smallo(L)\\
& \leq  & H(A_{[1:N]}^{[k]} | Q_{[1:N]}^{[k]}) - H(A_{\calE}^{[k]} | Q_{[1:N]}^{[k]})+ \smallo(L) \\
& \leq  & H(A_{[1:N]}^{[k]} | Q_{[1:N]}^{[k]}) - \frac{E}{N} H(A_{[1:N]}^{[k]} | Q_{[1:N]}^{[k]})+ \smallo(L) , \label{eql}
\end{eqnarray}
where $(a)$ follows from Lemma~\ref{thm:noQ} and (b) follows from 
the security constraint~\eqref{eqn:modelsecurity}. 
(\ref{eql}) follows by averaging over all $\calE$ with size $E$ and from Han's inequality.

%
%
%

Therefore, by letting $L \to \infty$,
$R = \frac{H(W_k)}{\sum_{n=1}^{N} H(A_n^{[k]})} \leq \frac{H(W_k)}{H(A_{[1:N]}^{[k]} | \queries)} \leq 1 - \frac{E}{N}$.

\subsection{The proof of $\rho \geq \frac{E}{N-E}$}
%
Note that \eqref{eqn:converserho} holds when $E \geq T$ as well. Then we have
$H(S) \geq  \frac{E}{N} \cdot H(A_{[1:N]}^{[k]} | Q_{[1:N]}^{[k]})$.
From (\ref{eql}), we have 
$H(A_{[1:N]}^{[k]} | Q_{[1:N]}^{[k]}) \geq \frac{N}{N-E} \left( H(W_k) - \smallo(L) \right)$. 
Hence, by letting $L \to \infty$, $\rho = \frac{H(S)}{H(W_k)} \geq \frac{E}{N-E} $.

\section{Conclusions}
We show that $C_{\text{ETPIR}} = \left( 1 - \frac{E}{N} \right) \cdot \left( 1 + \frac{T-E}{N-E} +  \left( \frac{T-E}{N-E} \right)^2 + \cdots +  \left( \frac{T-E}{N-E} \right)^{K-1} \right)^{-1}$ when $E < T$, and $\left( 1-\frac{E}{N} \right)$ when $E \geq T$. 
It is surprising that by adding eavesdroppers and allowing servers to collude in the PIR problem, there is a synergistic effect and the problem behaves differently in two regimes. In particular, when $E \geq T$, the PIR problem behaves like SPIR.
The ETPIR problem includes many previous works as special cases (refer to Figure~\ref{fig:intro_graph}), and the present work capitalizes on all ideas used in these works. Further, new challenges arise that go beyond previously studied problems, and new ideas are proposed to resolve the corresponding challenges. In particular, in the achievability proof when $E < T$ (Section~\ref{sec:schemeEsmallerT}), we require a layered MDS structure that simultaneously handles 3 correlated constraints -- security, privacy and correctness.

Regarding future work, it will be interesting to see whether including other elements would result in similar synergistic interactions, such as coded databases~\cite{banawan2018capacity,wang2017symmetric}, {\color{black} non-responsive servers~\cite{sun2017colluding}}, active adversaries~\cite{banawan2017capacity,wang2017secure,wang2018epsilon}, multi-round protocols~\cite{sun2018multiround}, arbitrary message length~\cite{sun2017arbitraryML, tian2018capacity} {\color{black}and secure storage~\cite{yang2018private, jia2018cross}} {\it etc}. 
Another promising direction is to find explicit precoding matrices when $E < T$ as we have only shown the existence in our schemes. 
In particular, it is interesting to see if Vandermonde matrices are sufficient. Also, our correctness conditions are over-constrained such that relaxations might lead to field size reduction.


\bibliographystyle{IEEEtran}
\bibliography{PIR}

\begin{thebibliography}{10}
\providecommand{\url}[1]{#1}
\csname url@samestyle\endcsname
\providecommand{\newblock}{\relax}
\providecommand{\bibinfo}[2]{#2}
\providecommand{\BIBentrySTDinterwordspacing}{\spaceskip=0pt\relax}
\providecommand{\BIBentryALTinterwordstretchfactor}{4}
\providecommand{\BIBentryALTinterwordspacing}{\spaceskip=\fontdimen2\font plus
\BIBentryALTinterwordstretchfactor\fontdimen3\font minus
  \fontdimen4\font\relax}
\providecommand{\BIBforeignlanguage}[2]{{%
\expandafter\ifx\csname l@#1\endcsname\relax
\typeout{** WARNING: IEEEtran.bst: No hyphenation pattern has been}%
\typeout{** loaded for the language `#1'. Using the pattern for}%
\typeout{** the default language instead.}%
\else
\language=\csname l@#1\endcsname
\fi
#2}}
\providecommand{\BIBdecl}{\relax}
\BIBdecl

\bibitem{sun2017capacity}
H.~Sun and S.~A. Jafar, ``The capacity of private information retrieval,''
  \emph{IEEE Transactions on Information Theory}, vol.~63, no.~7, pp.
  4075--4088, 2017.

\bibitem{sun2017colluding}
------, ``The capacity of robust private information retrieval with colluding
  databases,'' \emph{IEEE Transactions on Information Theory}, vol.~64, no.~4,
  pp. 2361--2370, 2018.

\bibitem{tajeddine2017private}
R.~Tajeddine, O.~W. Gnilke, D.~Karpuk, R.~Freij-Hollanti, C.~Hollanti, and
  S.~El~Rouayheb, ``Private information retrieval schemes for codec data with
  arbitrary collusion patterns,'' in \emph{IEEE International Symposium on
  Information Theory (ISIT)}.\hskip 1em plus 0.5em minus 0.4em\relax IEEE,
  2017, pp. 1908--1912.

\bibitem{jia2017capacity}
Z.~Jia, H.~Sun, and S.~A. Jafar, ``The capacity of private information
  retrieval with disjoint colluding sets,'' in \emph{GLOBECOM 2017-2017 IEEE
  Global Communications Conference}.\hskip 1em plus 0.5em minus 0.4em\relax
  IEEE, 2017, pp. 1--6.

\bibitem{banawan2017capacity}
K.~Banawan and S.~Ulukus, ``The capacity of private information retrieval from
  {B}yzantine and colluding databases,'' \emph{IEEE Transactions on Information
  Theory}, 2018.

\bibitem{wang2017EPIR}
Q.~Wang and M.~Skoglund, ``Secure private information retrieval from colluding
  databases with eavesdroppers,'' in \emph{2018 IEEE International Symposium on
  Information Theory (ISIT)}.\hskip 1em plus 0.5em minus 0.4em\relax IEEE,
  2018, pp. 2456--2460.

\bibitem{wang2018BEpir}
------, ``On {PIR} and symmetric {PIR} from colluding databases with
  adversaries and eavesdroppers,'' \emph{IEEE Transactions on Information
  Theory}, 2018.

\bibitem{banawan2018pir_wiretap}
K.~Banawan and S.~Ulukus, ``Private information retrieval through wiretap
  channel {II}: Privacy meets security,'' \emph{arXiv preprint
  arXiv:1801.06171}, 2018.

\bibitem{sun2016symmetric}
H.~Sun and S.~A. Jafar, ``The capacity of symmetric private information
  retrieval,'' \emph{IEEE Transactions on Information Theory}, 2018.

\bibitem{wang2017linear}
Q.~Wang and M.~Skoglund, ``Linear symmetric private information retrieval for
  {MDS} coded distributed storage with colluding servers,'' in
  \emph{Information Theory Workshop (ITW), 2017 IEEE}.\hskip 1em plus 0.5em
  minus 0.4em\relax IEEE, 2017, pp. 71--75.

\bibitem{wang2017secure}
------, ``Secure symmetric private information retrieval from colluding
  databases with adversaries,'' in \emph{Communication, Control, and Computing
  (Allerton), 2017 55th Annual Allerton Conference on}.\hskip 1em plus 0.5em
  minus 0.4em\relax IEEE, 2017, pp. 1083--1090.

\bibitem{schwartz1980fast}
J.~T. Schwartz, ``Fast probabilistic algorithms for verification of polynomial
  identities,'' \emph{Journal of the ACM (JACM)}, vol.~27, no.~4, pp. 701--717,
  1980.

\bibitem{zippel1979probabilistic}
R.~Zippel, ``Probabilistic algorithms for sparse polynomials,'' in
  \emph{Symbolic and algebraic computation}.\hskip 1em plus 0.5em minus
  0.4em\relax Springer, 1979, pp. 216--226.

\bibitem{cover2012elements}
T.~M. Cover and J.~A. Thomas, \emph{Elements of {I}nformation {T}heory}.\hskip
  1em plus 0.5em minus 0.4em\relax John Wiley \& Sons, 2012.

\bibitem{banawan2018capacity}
K.~Banawan and S.~Ulukus, ``The capacity of private information retrieval from
  coded databases,'' \emph{IEEE Transactions on Information Theory}, vol.~64,
  no.~3, pp. 1945--1956, 2018.

\bibitem{wang2017symmetric}
Q.~Wang and M.~Skoglund, ``Symmetric private information retrieval for {MDS}
  coded distributed storage,'' in \emph{IEEE International Conference on
  Communications (ICC)}.\hskip 1em plus 0.5em minus 0.4em\relax IEEE, 2017, pp.
  1--6.

\bibitem{wang2018epsilon}
Q.~Wang, H.~Sun, and M.~Skoglund, ``The $\epsilon $-error capacity of symmetric
  {PIR} with {B}yzantine adversaries,'' \emph{arXiv preprint arXiv:1809.03988},
  2018.

\bibitem{sun2018multiround}
H.~Sun and S.~A. Jafar, ``Multiround private information retrieval: Capacity
  and storage overhead,'' \emph{IEEE Transactions on Information Theory}, 2018.

\bibitem{sun2017arbitraryML}
------, ``Optimal download cost of private information retrieval for arbitrary
  message length,'' \emph{IEEE Transactions on Information Forensics and
  Security}, vol.~12, no.~12, pp. 2920--2932, 2017.

\bibitem{tian2018capacity}
C.~Tian, H.~Sun, and J.~Chen, ``Capacity-achieving private information
  retrieval codes with optimal message size and upload cost,'' \emph{arXiv
  preprint arXiv:1808.07536}, 2018.

\bibitem{yang2018private}
H.~Yang, W.~Shin, and J.~Lee, ``Private information retrieval for secure
  distributed storage systems,'' \emph{IEEE Transactions on Information
  Forensics and Security}, vol.~13, no.~12, pp. 2953--2964, 2018.

\bibitem{jia2018cross}
Z.~Jia, H.~Sun, and S.~A. Jafar, ``Cross subspace alignment and the asymptotic
  capacity of $ x $-secure $ t $-private information retrieval,'' \emph{arXiv
  preprint arXiv:1808.07457}, 2018.

\end{thebibliography}
\end{document}